\documentclass[prd,8pt,aps,showpacs,twocolumn,superscriptaddress,nofootinbib,notitlepage,floatfix]{revtex4-1}

\RequirePackage{mathrsfs}

\def\eq#1{{Eq.~(\ref{#1})}}
\newcommand{\Le}{\left(}
\newcommand{\Ra}{\right)}

\newcommand{\beq}{\begin{equation}}
\newcommand{\eeq}{\end{equation}}
\newcommand{\beqar}{\begin{eqnarray}}
\newcommand{\eeqar}{\end{eqnarray}}
\newcommand{\caB}{{\cal B}}

\usepackage{color}
\newcommand{\revisionZ}[1]{{#1}}
\newcommand{\revision}[1]{{#1}}
\newcommand{\revisionF}[1]{{#1}}
\newcommand{\revisionFF}[1]{{#1}}
\newcommand{\revisionU}[1]{{#1}}

\begin{document}
\title{ The dimensionally reduced description of the high energy scattering and the effective action for the reggeized gluons}

\author{S.~Bondarenko}
\affiliation{Physics Department, Ariel University, Ariel 40700, Israel}

\author{M.A.Zubkov 
\thanks{on leave of absence from Institute for Theoretical and Experimental Physics, B. Cheremushkinskaya 25, Moscow, 117259, Russia}%
}                     
\affiliation{Physics Department, Ariel University, Ariel 40700, Israel}

%
%
%
%
\begin{abstract}
We discuss the high energy scattering of hadrons with the multi - Regge kinematics. The effective $2D$ model for the interaction between the produced hard gluons appears as a result of the dimensional reduction, which is similar to the dimensional reduction $3+1$ D $\to 3$ D of the high temperature gauge theory. It is demonstrated that being supplemented by the exchange by virtual $3+1$ D gluons this $2D$ model gives rise to the hermitian version of the effective action of Lipatov for the interaction between the ordinary and the reggeized gluons.
\end{abstract}
\maketitle

\section{Introduction}

The description of the hadron - hadron scattering at large energies $\sqrt{s}$ compared to the momentum transfer $\sqrt{t}$ is dominated by the amplitudes with the so - called multi - Regge kinematics \cite{EffAct}. The most probable process is the multi-gluon production. Let us denote the momenta of the two initial partons by $p_A$ and $p_B$. The final state gluon momenta are $k_0 = p_A^\prime , k_1, ..., k_n, k_{n+1} = p_B^\prime$. For such processes we have
\begin{equation}
s \gg  s_i = 2k_{i-1}k_i \gg t_i = q^2_i = \Big(p_A - \sum_{r=0}^{i-1}k_r\Big)^2\nonumber
\end{equation}
In the leading logarithmic approximation  the multi - gluon production amplitude has the form:
\begin{equation}
A_{LLA} \sim \Pi_{i=1}^{n+1} s_i^{\omega(t_i)}\label{0r}
\end{equation}
Here
\begin{equation}
\omega(t) = \frac{g^2}{(2\pi)^3}\frac{N_c}{2}\int \frac{d^2 k_\bot (-q^2_\bot)}{k_\bot^2  (k_\bot - q_\bot)^2}\label{1r}
\end{equation}
The infrared divergencies in $\omega$ cancel in the total cross section with
the contribution of the real gluons.

The generalization of this description may be given in terms of the effective field theory. The phenomenological Regge field theory based on the pomeron degrees of freedom was introduced in \cite{Gribov} (see also \cite{OldRFT,NewRFT,Odd}). The more refined theory deals with the dynamics of the colored reggeized gluons that interact with the ordinary virtual and real gluons. The majority of the  produced in the mentioned above process (with the multi - Regge kinematics) particles are gluons, which give rise to jets and are splitted into the partons of the created hadrons. \revisionF{Those gluons (we call them physical) eject and absorb virtual soft gluons with the momentum transfer much smaller than the momentum of the final state gluons. The collective states of the soft gluons called reggeons provide the interaction between the clasters of the hard gluons, these clusters are significantly separated in rapidity. At the same time there are virtual gluons ejected and absorbed by the gluons
in the same cluster and typically the parameter $\eta$ is introduced as an ultraviolet cutoff for the cluster's gluons relative rapidities.
Also, the rapidity parameter $\eta$ plays the role of the infrared cutoff for the interactions of the clusters by exchange of the reggeons.
One can say, that those groups of hard gluons, separated significantly in rapidity, emit and absorb the reggeons.}

The mentioned above effective theory of reggeized gluons was formulated by L.Lipatov \cite{Lipatov,LipatovEff}. \revisionF{It corresponds to the interactions between the physical gluons  in the multi - Regge kinematics.} In the center of inertia (of the two colliding hadrons) reference frame\footnote{\revisionF{For the two colliding hadrons we may call this reference frame the "center of mass" reference frame. However, in the following we will also speak of the center of inertia reference frame for the physical gluons. They  are massless. Therefore, we feel this appropriate to refer to this reference frame as to the center of inertia reference frame. By definition for the two colliding particles this is the reference frame, where the space component of the total four - momentum vanishes.}} the colliding partons  move along the parallel straight lines and have momenta close to the light cone. This distinguishes the direction in space $\vec{n}$ and determines the corresponding light cone coordinates\footnote{\revisionF{Vector $\vec{n}$ is directed along the $3$ - momenta of the two colliding partons in their center of inertia reference frame. If the reference frame is chosen in such a way that $\vec{n}$ is directed along the $z$ - axis, then the light cone coordinates are $\frac{t\pm z}{\sqrt{2}}$.}}.  In the effective theory of Lipatov in addition to the physical gluon field $A$ the so - called reggeon field $\cal B$ is added. The effective action describes their interaction. The main advantage of this theory is that instead of summing a lot of diagrams with the usual virtual gluons one may sum sufficiently smaller number of diagram that involve reggeons.

This effective theory of Lipatov \cite{EffAct}, in particular,  can be used for the calculation of both leading contributions and the sub-leading  corrections to the amplitudes and production vertices \cite{LipatovEff,EffAct,BKP,GLR,BK}. The investigation of this effective theory includes the solution of classical equations of motion in terms of the reggeon fields \cite{Our1,Bondarenko:2017vfc}. Inserting these solutions to the Lagrangian it is possible to develop the field theory in terms of the reggeon fields only (with the loop corrections taken into account). It is expected, that the results obtained in this way should \revision{generalize} the ones obtained via the small-x BFKL approach \cite{BFKL,BFKLInt}. Recently it has been checked, for example, that the gluon Regge trajectory, which determines the form of the propagator of the reggeized gluon obtained within the framework of the Lipatov effective theory corresponds to Eq. (\ref{1r}). \revisionF{It is worth mentioning that the gluon Regge trajectory has been calculated up to two loops from the Lipatov's effective action in  \cite{RR1}, while the triple Pomeron vertex has been re - obtained from the high energy effective action in \cite{RR2} (the triple Pomeron vertex \cite{TripleV} is, actually, the interaction vertex of six reggeon fields). It would also be interesting to calculate the BFKL kernel, which is an effective vertex of  the interaction between four reggeons.} Another parallel approach is the theory of the Color Glass Condensate \cite{Venug,Kovner,Hatta1}, which operates with the semi - phenomenological notion of the color density function. In principle, the effective theory of Lipatov should be able to reproduce the basic features of this approach as well as of the other formalisms related to the resummation of leading logarithms and to the description of the processes with the multi - Regge kinematics associated with the  small-x physics. Among the latter approaches we notice the dipole approximation \cite{3}, the Balitsky -- Kovchegov (BK) equation \cite{4,5}, the Balitsky -- Jalilian -- Marian -- Iancu -- McLerran --
Weigert -- Leonidov -- Kovner (BJIMWLK) equation \cite{6,7}, and the calculus of the  Wilson lines \cite{8,9,10,4,13,14}. The work in the mentioned above adjacent directions is carried out currently by several theoretical groups.

Originally L.Lipatov came to his action on the basis of the analysis of the diagram resummation in the perturbative QCD. In the present paper we propose the completely different and a relatively simple way to derive the reggeon theory. As a result we will obtain the extension of the action of Lipatov. Our derivation is based on the combination of the effective $2D$ model for the high energy scattering with the perturbative exchange by virtual gluons.

The light cone coordinates of the original version of Lipatov theory correspond to the direction of motion of the colliding hadrons in their center of inertia reference frame.
However, we may also apply the same approach to the $i$ - the sector of the multi - Regge kinematics with the produced particles (physical gluons) that have close values of rapidities. Because $s_i \gg t_i$, in order to calculate the leading contribution of the interaction between the $i-1$ - th and the $i$ - th  produced particles, we may imagine that those two particles existed forever, that they collide instead of the partons of the originally colliding hadrons. We may come to the reference frame, where the total $3$ momentum of the pair is zero. This fixes the new light - cone coordinates. Our approach allows to derive the effective reggeon action in these coordinates. Thus we come to the version of Lipatov effective theory that describes the $i$ - th sector of the scattering with the multi - Regge - kinematics. Using the Eikonal approximation for the multi - gluon ejection/absorbtion by the high energy particles, we are able to express in this effective theory the contribution to the scattering amplitude through the propagator of the reggeon.

In QCD the derivation of the expression for the radiation of multiple virtual gluons by physical gluon with high energy in the Eikonal approximation has been given in many publications (see, for example, \cite{Eikonal}). This approximation is relevant for the multi - Regge kinematics because it is valid when the physical gluon momentum is changed slightly compared to itself. The result is that the scattering amplitude of particles is given by the correlator of the Wilson lines taken along the naive trajectories of the scattered particles. The Wilson lines entering this expression are to be taken in the adjoint representation for the interaction between the physical gluons. On the other hand, under these conditions the leading contribution to the multi - gluon exchange between the physical particles is described by the exchange via the so - called reggeized gluon.

In \cite{Verlinde} E. Verlinde and H. Verlinde have proposed the $2D$ effective action for the high - energy scattering. Later a version of this action was derived by Aref'eva using the technique of lattice gauge theory \cite{Arefeva}. The output of the derivation is the similarity between the high - energy scattering with the Regge kinematics and the high temperature gauge theory. In the finite temperature theory because of the dimensional reduction the theory becomes effectively three - dimensional. In the description of the processes with Regge kinematics the two dimensions corresponding to the light cone coordinates $x^\pm$ are reduced, and the resulting effective theory is two - dimensional.
Here we do not follow the derivation of \cite{Verlinde,Arefeva}. Instead we consider the dimensional reduction $3+1$ D $\to$ $2$ D on the same grounds as the dimensional reduction is done usually within the finite temperature gauge theory. Namely, we may simply consider the region of space - time, in which the coordinates $x^\pm$ belong to small intervals while gauge fields do not depend on $x^\pm$. Qualitatively the reason for this reduction is clear - when we describe the high energy scattering in the center of inertia reference frame, the scattered particles have such a large energy that during the collision the fields $A^\pm,A_a$ have not time to change. The resulting continuum $2D$ action will differ somehow from that of \cite{Verlinde,Arefeva}.

The effective action, which will be obtained, contains the two dimensionless parameters. It operates with the Wilson lines
 \begin{eqnarray}
{\cal T}_+(x_\bot) &=& P\, {\rm exp}\,\Big(\pm i \int_{-\infty}^{\infty} A^-(\bar{x}^+,x^-,\vec{x}_\bot)d \bar{x}^+ \Big)_{x^-=0} \nonumber\\
{\cal T}_-(x_\bot) &=&P\, {\rm exp}\,\Big(\pm i \int_{-\infty}^{\infty} A^+(x^+,\bar{x}^-,\vec{x}_\bot)d\bar{x}^- \Big)_{x^+=0}
\end{eqnarray}
that depend on the coordinates $\vec{x}_\bot$ in the plane orthogonal to vector $\vec{n}$ that marks the direction, along which the scattered particles move in the center of inertia reference frame. The mentioned effective action describes bare interaction between the two high energy physical gluons. The corrections to this bare interaction may be taken into account if we simply add the exchange by the virtual gluons described by the ordinary action of gluodynamics. In the framework of the $3+1$ D theory the interaction of $2D$ model between $\cal T$ are global while the exchange by virtual gluons is local. This allows to avoid the overcounting.

The paper is organized as follows. In Section \ref{2D} we describe the dimensional reduction process that gives the effective $2D$ model. The output of this section is the effective action that contains two dimensionless parameters. In Section \ref{4D} we
remind the reader the description of the multi - gluon ejection/absorbtion in Eikonal approximation. In Sect. \ref{Lipatov} we demonstrate how the effective reggeon field may be introduced into the theory with the action that contains both the effective $2D$ model part and the part corresponding to the exchange by virtual gluons. In Sect. \ref{Lipatov2} we show how the effective action of Lipatov appears for the particular choice of the parameters of the model. In Sect. \ref{sector} we apply the same construction to the $i$ - th sector of the hadron - hadron scattering with the multi - Regge kinematics. In Sect. \ref{concl} we end with the conclusions.

\section{The effective $2D$ model for the high energy scattering}

\label{2D}

  In continuum theory we may chose the following dynamical variables.
The $2D$ components of the gauge field $A_i(x^+,x^-,\vec{x}_\bot)$ $\in$ $su(3)$ ($i=2,3$) and the extra $su(3)$ fields $A_+(x^+,x^-,\vec{x}_\bot)$, $A_-(x^+,x^-,\vec{x}_\bot)$ that originate from the light cone components of the original $4D$ gauge field.
We start from the pure gauge field action
$$
S=-\frac{1}{2 g^2}{\rm Tr}\,\int d^4 x \, G_{\mu\nu}^2
$$
where $G_{\mu \nu}$ is the field strength while $g$ is the coupling constant. We chose the $4$ - vector  $n^\mu = (1,1,0,0)$ and denote $\vec{n} = (1,0,0)$. The light - cone notations are:
$$
x^- = x_+ = \frac{x^0 - x^1}{\sqrt{2}}, \quad x^+ = x_- = \frac{x^0 + x^1}{\sqrt{2}},$$ $$ \vec{x}_\bot = (0,0,x^2,x^3)
$$
We rewrite the action in the light - cone coordinates as follows
\begin{eqnarray}
S&=&-\frac{1}{2 g^2}{\rm Tr}\,\int d^4 x \, G_{ij}G^{ij} -\frac{1}{2 g^2}{\rm Tr}\,\int d^4 x \, G_{\alpha\beta}G^{\alpha\beta}\nonumber\\&&-\frac{1}{ g^2}{\rm Tr}\,\int d^4 x \, G_{\alpha i}G^{\alpha i} \label{Gs}\\&& \quad i,j = 2,3; \alpha,\beta = \pm\nonumber
\end{eqnarray}

   The direct dimensional reduction $3+1$D $\to 2$D works as follows. We should consider the $3+1$ D space - time, in which the coordinates $x^\pm$ belong to the interval $0<x^\pm<\Delta$, where $\Delta = \frac{1}{\Lambda}$ and $\Lambda$ is the parameter of the dimension of mass, which is supposed to be much larger than the typical energies of the virtual gluons radiated and absorbed during the scattering process.
  \revisionF{The integration over $x^\pm$ is effectively reduced to the integration over the intervals of lengths $1/\Lambda$ because we describe the high energy scattering in the center of inertia reference frame, and the scattered particles have such a large energy that during the collision the fields $A^\pm,A_a$ have not time to change. $1/\Lambda$ is of the order of the collision time. The uncertainty relation assumes $\Lambda \sim \sqrt{s}$. The discussion of the  different approaches to the description of this dimensional reduction may be found in \cite{Verlinde,Arefeva}.}

   Because of the large value of $\Lambda$ we neglect the dependence of $A$ on $x^\pm$. As a result Eq. (\ref{Gs}) receives the form:
\begin{eqnarray}
S&=&-\frac{1}{2 \Lambda^2 g^2}{\rm Tr}\,\int d^2 \vec{x}_\bot \, G_{ij}G^{ij} +\frac{1}{\Lambda^2 g^2}{\rm Tr}\,\int d^4 x \, [A_+,A_-]^2\nonumber\\&&+\frac{2}{ g^2}{\rm Tr}\,\int d^4 x \, (D_{i}A_+)(D_i A_-) \label{Gs2}
\end{eqnarray}
The value of $\Lambda$ is of the order of magnitude of $\sqrt{s}$ while the typical momentum transfer is $\sqrt{-t} \ll \sqrt{s}$. Therefore, we neglect the terms containing $\Lambda^2$ in the denominator. As a result we are left with the action
\begin{eqnarray}
\revisionU{S_{+-}}&=&\frac{2}{ g^2}{\rm Tr}\,\int d^4 x \, (D_{i}A_+)(D_i A_-) \label{Gs3}
\end{eqnarray}
where $D_i$ is the covariant derivative with respect to the $2D$ gauge field.

   Instead of the algebra elements $A_\pm$ it will be more useful to operate with the group elements of the original $4D$ theory. We will be interested in the following in the adjoint Wilson lines:
\revisionF{\begin{eqnarray}
{\cal T}_+(x_\bot) &=&  {\rm exp}\,\Big( i \int_{0}^{\Delta} A^-(\vec{x}_\bot)d \bar{x}^+ \Big) \nonumber\\
{\cal T}_-(x_\bot) &=& {\rm exp}\,\Big( i \int_{0}^{\Delta} A^+(\vec{x}_\bot)d\bar{x}^- \Big)\label{E8}
\end{eqnarray}
Here the integral in the exponent is trivial because in the discussed effective $2D$ theory the field $A^\pm$ does not depend on $x^\pm$. Therefore, the path ordering is not needed.} We denote
$$
q^- = q_+ = \frac{q_0 + q_1}{\sqrt{2}}= \frac{q^0 - q^1}{\sqrt{2}}$$ $$ q^+ = q_- = \frac{q_0 - q_1}{\sqrt{2}}=\frac{q^0 + q^1}{\sqrt{2}}$$
$$ \vec{q}_\bot = (0,0,q_3,q_4)
$$
With these notations
$$
q_i x^i = q^0x^0 - \vec{q}\vec{x} = q_+ x^+ + q_- x^- - (\vec{q}_\bot \vec{x}_\bot)
$$
Inside the effective dimensionally reduced $2D$ theory the effective action Eq. (\ref{Gs3}) operates with the field $A$ that does not depend on $x^\pm$.
\revisionU{The proposed dimensional reduction assumes the smallness of the product $A_\pm \Delta$, therefore Eq. (\ref{Gs3}) may be rewritten with the help of  ${\cal T}_\pm,{\cal T}_\pm^+$ as follows}
\begin{eqnarray}
S_{+-} &=&\revisionU{-} \int d^2 \vec{x}_\bot \, \sum_{a=2,3}\,{\rm Tr}\,\Big(\zeta\,D_a {\cal T}_+ D_a {\cal T}_- + \zeta D_a {\cal T}^\dagger_+ D_a {\cal T}^\dagger_-\nonumber\\&& - \zeta^\prime D_a {\cal T}_+ D_a {\cal T}^\dagger_- - \zeta^\prime D_a {\cal T}^\dagger_+ D_a {\cal T}_-\Big)\label{S+-}
\end{eqnarray}
Here the constants $\zeta, \zeta^\prime$ obey
\begin{equation}
\zeta + \zeta^\prime = \frac{1}{g^2}\label{naive0}
\end{equation}
i.e.
\begin{equation}
\zeta =  \frac{1}{2g^2} + \frac{\Delta\zeta}{2}, \quad \zeta^\prime = \frac{1}{2g^2}-\frac{\Delta\zeta}{2}\label{naive00}
\end{equation}
\revisionU{The equivalence between \eq{S+-} and \eq{Gs3}, when $A$ does not depend on $x^\pm$ while $A_\pm \Delta \ll 1$, can be demonstrated by the following calculation. We have
for the first term of \eq{S+-} to leading order precision:
\begin{eqnarray}
&&  \int d^2 \vec{x}_\bot \, {\rm Tr}\, D_a {\cal T}_+ D_a  {\cal T}_-  \nonumber\\&&=\int d^2 \vec{x}_\bot \, {\rm Tr}\,D_a e^{ i \int_{0}^{\Delta} A^-(\vec{x}_\bot)d {x}^+ } D_a e^{ i \int_{0}^{\Delta} A^+(\vec{x}_\bot)d {x}^- }\nonumber\\&&=-\int d^2 \vec{x}_\bot \,\int_{0}^{\Delta} d {x}^+  \int_{0}^{\Delta} d {x}^- {\rm Tr}\,D_a    A^-(\vec{x}_\bot)  D_a   A^+(\vec{x}_\bot)
\nonumber\\&&=-\int d^2 \vec{x}_\bot  d {x}^+  d {x}^- {\rm Tr}\,D_a    A^-  D_a   A^+
\end{eqnarray}
Calculating similarly the other terms of \eq{S+-} and summing all terms together we will obtain finally:
\begin{eqnarray}
S_{+-} &=&\revisionU{-} \int d^2 \vec{x}_\bot \, \sum_{a=2,3}\,{\rm Tr}\,\Big(\zeta\,D_a {\cal T}_+ D_a {\cal T}_- + \zeta D_a {\cal T}^\dagger_+ D_a {\cal T}^\dagger_-\nonumber\\&& - \zeta^\prime D_a {\cal T}_+ D_a {\cal T}^+_- - \zeta^\prime D_a {\cal T}^\dagger_+ D_a {\cal T}_-\Big)\nonumber\\
&=&2\int d^4 x \, \sum_{a=2,3}\,{\rm Tr}\,D_a    A^-  D_a   A^+ \Big(\zeta+ \zeta^\prime\Big)
\end{eqnarray}
and since $\zeta + \zeta^\prime = \frac{1}{g^2}$ we indeed reproduce  Eq. (\ref{Gs3}).}

\revision{With the definition}
$$
{\cal T}_0 = \frac{1}{\sqrt{2}}({\cal T}_+ + {\cal T}_-), \quad {\cal T}_1 = \frac{1}{\sqrt{2}}({\cal T}_+ - {\cal T}_-)$$$$ {\cal T}^0 = {\cal T}_0, \quad {\cal T}^1 = -{\cal T}_1
$$
$S_{+-}$ receives the form:
\begin{eqnarray}
S_{+-} &=& \revisionU{-}\int d^2 \vec{x}_\bot \, \sum_{I = 0,1}\sum_{a=2,3}\,\frac{1}{2}{\rm Tr}\,\Big(\zeta\,D_a {\cal T}_I D_a {\cal T}^I \nonumber\\&&+ \zeta D_a {\cal T}^\dagger_I [D_a {\cal T}^I]^\dagger- 2 \zeta^\prime D_a {\cal T}^I D_a {\cal T}^\dagger_I  \Big)\label{S+-2}
\end{eqnarray}

Since this form of the action does not contain dimensional parameters it may be used in the framework of the $3+1$ D theory for the bare description of the high energy scattering at $s/(-t) \gg 1$, and by $\cal T$ we should understand
\begin{eqnarray}
{\cal T}_+(x_\bot) &=& P\, {\rm exp}\,\Big(\pm i \int_{-\infty}^{\infty} A^-(\bar{x}^+,x^-,\vec{x}_\bot)d \bar{x}^+ \Big)_{x^-=0} \nonumber\\
{\cal T}_-(x_\bot) &=&P\, {\rm exp}\,\Big(\pm i \int_{-\infty}^{\infty} A^+(x^+,\bar{x}^-,\vec{x}_\bot)d\bar{x}^- \Big)_{x^+=0}\label{E13}
\end{eqnarray}

\revisionF{This is the basic idea of the present paper: we consider the $3+1$ D theory, and take as a first approximation to the effective action the  action of the reduced $2D$ theory Eq. (\ref{S+-2}), where instead of the group elements of Eq. (\ref{E8}) the Wilson lines of Eq. (\ref{E13}) are substituted. Next, we take into account an extra exchange by the virtual soft gluons and thus come to the effective action of Lipatov.}

We will see in the next sections that for the simplest choice of parameters with
\begin{equation}
\zeta = \zeta^\prime = \frac{1}{2g^2}\label{naive}
\end{equation}
the \revisionZ{hermitian version \cite{Nefedov} of the} effective action of Lipatov is reproduced.
Those values may be considered as the zero approximation to $\zeta$ and $\zeta^\prime$. We expect that in the more refined theory the values of $\zeta$ and $\zeta^\prime$ deviate from Eq. (\ref{naive}) and should be taken from experiment.
For the case $\Delta \zeta = 0$ we may rewrite the $2D$ action as follows
\begin{eqnarray}
S_{+-} &=& \revisionU{-} \int d^2 \vec{x}_\bot \, \sum_{I = 0,1}\sum_{a=2,3}\,\frac{1}{4g^2}{\rm Tr}\,D_a ({\cal T}^I-[{\cal T}^I]^\dagger)\label{S+-3}\\&& D_a ({\cal T}_I-{\cal T}_I^\dagger)\nonumber\\&=& \revisionU{-}\int d^2 \vec{x}_\bot \, \sum_{a=2,3}\,\frac{1}{2g^2}{\rm Tr}\,D_a ({\cal T}_+-{\cal T}_+^\dagger) D_a ({\cal T}_--{\cal T}_-^\dagger)\nonumber
\end{eqnarray}
\revisionF{In this case the action is consistent with the effective Lipatov gluon production vertex\footnote{\revisionF{The discussion of various effective vertices in the Lipatov effective action approach may be found in \cite{BP2018} and references therein.}}. We do not exclude, that the formal leading order perturbative description of the multi - Regge kinematics with the modified gluon production vertex is consistent with the effective field theory corresponding to the more general choice of couplings of the form of Eq. (\ref{naive00}) with nonzero $\Delta \zeta$. The consideration of this issue is, however, out of the scope of the present paper.}

\section{Radiation of multiple virtual gluons in the eikonal approximation}
\label{4D}

Here we remind the reader the basic facts about the eikonal approximation \cite{Eikonal}.
In this section we consider the scattering process of several high energy physical particles (gluons).  We consider the situation, when the momentum of this particle remains close to its initial value. This is the case, when the momentum transfer is small compared to the total momentum. Then the so - called eikonal approximation may be applied to the description of scattering. Such a particle radiates multiple virtual gluons, which, in turn, interact with each other. Let us consider contribution to the scattering amplitude of the external physical gluon that enters the scattering process in the state described by a vector in color space $\epsilon(p_i)$ and quits scattering with $\bar{\epsilon}(p_f)$, where $p_f \approx p_i$. We assume, that those two states are connected by the continuous worldline, i.e. the outgoing gluon is not created during the scattering. The overall amplitude is equal to the integral over the gauge field $A$ of the product of the quantity $M_{f i}(A)$ (to be defined below) and the similar quantities corresponding to the other colliding particles.

The direct calculation gives the contribution to the scattering amplitude  for the radiation of multiple virtual gluons by a physical gluon \cite{Eikonal}:
\begin{eqnarray}\label{SSec741}
M_{f i}(A)\,
&=&\,\,\frac{2E}{Z_g}\,
\int\,d^{3}\vec{x}\,e^{-\imath\,\Le\,\vec{q}\vec{x} \Ra}\,\bar{\epsilon}(p_f)\nonumber\\&&\Le\,P\,e^{\imath \,\int_{-\infty}^{\infty}\,dt\, n^{\mu}\,A_{\mu}(t,\vec{x}+\vec{n}t)\,}\,-\,1\,\Ra\,{\epsilon}(p_i)
\end{eqnarray}
where the P - ordered exponent is taken in the adjoint representation. Here $Z_g$ is the gluon wave function renormalization constant. Four vector $n^\nu$ marks the trajectory of the particle.

On the basis of Eq. (\ref{SSec741}) we come to the conclusion that the overall scattering amplitude of the two high energy gluons in the center of inertia reference frame with the momentum transfer $q = (0,\vec{q})$ is equal to
\begin{eqnarray}
&&\frac{2s}{Z_g^2} D(\vec{q})(2\pi)^4 \delta^{(4)}(0)
= \frac{2s}{Z_g^2}\frac{1}{Z^\prime}\int DA  e^{i S_{}[A]} \nonumber\\&&\,
\int\,d^{3}x\,e^{-\imath\,\Le\,\vec{q}\vec{x}\Ra}\,\Le\,P\,e^{\imath \,\int_{-\infty}^{\infty}\,dt  n^{\mu}\,A_{\mu}(t,\vec{x}+\vec{n}t)\,}\,-\,1\,\Ra\, \,\nonumber\\&&\otimes
\int\,d^{3}y\,e^{\imath\,\Le\,\vec{q}\vec{y}\Ra}\,\Le\,P\,e^{\imath \,\int_{-\infty}^{\infty}\,dt \, n^{\prime \mu}\,A_{\mu}(t,\vec{y}+\vec{n}^\prime t)\,}\,-\,1\,\Ra\,\nonumber\\
&&Z^\prime = \int DA  e^{i S_{}[A]}, \quad n^\prime = (1,\vec{n}^\prime) = (1,-\vec{n})
\label{Dq0}
\end{eqnarray}
The kinematics implies that the momentum transfer $q$ has the components $q^0=0$ and $\vec{q} = \vec{q}_\bot \bot \vec{n}$.
The divergent factor $\delta^4(0)$ on the left hand side of Eq. (\ref{Dq0}) corresponds to the $4$ - momentum conservation, which leads to the same factor on the right hand side.
Recall, that the P - exponent is taken in the adjoint representation. Function $D$ may be identified with the propagator of a multi - gluon composite object. We will see below in Sect. \ref{sector}, that it may be expressed also through the propagator of the reggeized gluon (reggeon).

Let us use the fact that for the high energy particles moving in opposite directions the interactions occur only in a small vicinity of the moment, when they meet each other. \revisionF{The colliding particles move along vector $\vec{n}$ in the opposite directions with the velocity of light. They meet together when the projections of their coordinates to $\vec{n}$ coincide. At this moment the interaction occurs. The interaction assumes slight change of momentum, and this change of momentum in the Eikonal approximation is neglected at all. Therefore, we may think that those particles interact only in the moment of meeting when the projection to $\vec{n}$ of their coordinates coincide.
That is, we should take $(\vec{x} - \vec{y})\vec{n} = 0$.}
Then we come to
\begin{eqnarray}
{ D}(q_\bot)&=&\frac{1}{ Z^\prime }\int DA  e^{i S_{}[A]} \,
\int \,d^{2}\vec{x}_\bot\,e^{-\imath\,\Le\,\vec{q}_\bot\vec{x}_\bot\Ra}\nonumber\\&&\, \Le\,P\,e^{\imath \,\int_{-\infty}^{\infty}\,d t\, (A^0(t, \vec{x}_\bot + \vec{n}t) - \vec{n}\,\vec{A}(t, \vec{x}_\bot + \vec{n}t))\,}\,-\,1\,\Ra\, \,\nonumber\\&&\otimes
\,\Le\,P\,e^{\imath \,\int_{-\infty}^{\infty}\,d t\, (A^0(t, -\vec{n}t) + \vec{n}\,\vec{A}(t, - \vec{n}t))\,}\,-\,1\,\Ra\,\label{Dq2}
\end{eqnarray}

\section{Effective action for the reggeized gluons}
\label{Lipatov}

Below we will show, that the effective action proposed by Lipatov (see, for example,  Eq. (210) of \cite{Lipatov}) appears, when we consider the ordinary QCD action $S^{0}[A]$ supplemented by the $2D$ model introduced above. We justify this procedure as follows. Bare effective $2D$ action for the high energy scattering is the action of the deduced above model. The correlations between the fields ${\cal T}_{\pm}$ in this model are obtained as a result of the interaction with the color gauge field in the approximation when it does not vary along the light cone. This may be considered as the  exchange by bare reggeon. To obtain corrections to this exchange we should take into account the interactions with the ordinary virtual gluons. In order to take into account these corrections we add the action of the pure QCD $S^{(0)}[A]$ to the action of the $2D$  model of Eq. (\ref{S+-}). At the same time in the latter we substitute the covariant derivative $D_i$ by the ordinary one, which means that we fix the boundary conditions with the vanishing values of  $A_i(\pm \infty, x^-, \vec{x}_\bot)$ and $A_i(x^+,\pm \infty, \vec{x}_\bot)$ for $i = 2,3$.

  Since the effective two - dimensional model appears in the dimensional reduction of gluodynamics, it is already included into the action $S^{(0)}$.
Therefore, we cannot simply add one to another. We should add an additional prescription, which allows to avoid overcounting in the further perturbative calculations. The prescription for this avoiding is actually very simple. Its essence is the understanding that the reggeon fields appear as the classical solutions of the equations of motion of gluodynamics. Namely, we will see that the theory with the action that consists of Eq. (\ref{S+-}) and $S^{(0)}$ may be rewritten in terms of the auxiliary field $\cal A$ that plays the role of the reggeon field. In the presence of this reggeon field the classical solution for the ordinary gluon field $A$ gives $A_\pm = {\cal A}_\pm$. Thus, the \revisionZ{most natural way to construct the perturbation theory assumes that the perturbations around this classical solution are considered. However, this procedure would lead to the double counting of the reggeons. Therefore, in order to avoid the overcounting we should build the perturbation theory around $A=0$.} We will see, that this corresponds to the choice of the initial parameters of the effective reggeon theory proposed by L.N.Lipatov in the most recent publications on this subject \cite{Lipatoveff2}. \revisionZ{At the same time the perturbation theory built in this way is identical precisely to that of based on the expansion around $A_\pm = {\cal A}_\pm$ in the theory with the modified coefficient in front of the kinetic term of $\cal A$. Written in this form the coefficients in the effective reggeon theory match another conventions assumed, for example, in Eq. (210) of \cite{Lipatov}.}

We consider the two sets of variables: the non - local $2D$ fields ${\cal T}_\pm (x_\bot)$ and the local gluon fields $A(x^+,x^-,x_\bot)$. We suppose, that the local fields tend to the trivial values at infinity.
Thus we consider QCD with the following modified action
\begin{eqnarray}
S[A] & = &S^{0}[A] \revisionU{-}  \int d^2 x_\bot \, {\rm Tr} \, \Big(\zeta\,\partial_a {\cal T}_+ \partial_a {\cal T}_- + \zeta \partial_a {\cal T}^\dagger_+ \partial_a {\cal T}^\dagger_-\nonumber\\&& - \zeta^\prime \partial_a {\cal T}_+ \partial_a {\cal T}^\dagger_- - \zeta^\prime \partial_a {\cal T}^\dagger_+ \partial_a {\cal T}_-\Big)
\end{eqnarray}
Let us denote $$\zeta_{AB} = \left(\begin{array}{cc}\zeta & -\zeta^\prime \\ -\zeta^\prime & \zeta \end{array}\right)$$
In the following we will use matrices $\lambda^a$ that are the generators of the $U(3)$ group in the adjoint representation. (The matrix proportional to unity is added to the set of the generators of $SU(3)$.) We normalize those matrices of the generators of the $U(3)$ group in adjoint representation in such a way that ${\rm Tr}\,\lambda^a\,\lambda^b = \delta^{ab}$. Next, we introduce the new variable ${\cal A}^\pm={\cal A}_\mp = {\cal A}_\mp^a \lambda^a \in u(3)\oplus u(3)$. Here the components ${\cal A}_\mp^a$ are the complex numbers.

We introduce the following notation
\begin{eqnarray}
{\cal T}_+(x^-,x_\bot) &=&P\, {\rm exp}\,\Big( i \int A^-(x^+,x^-,\vec{x}_\bot)d x^+ \Big)_{adj}\in u(3)\nonumber\\
{\cal T}_-(x^+,x_\bot) &=& P\, {\rm exp}\,\Big( i \int A^+(x^+,x^-,\vec{x}_\bot)dx^- \Big)_{adj}\in u(3)\nonumber\\&&\label{TT}
\end{eqnarray}
for the adjoint Wilson line. In the following we assume, that all Wilson lines are taken in the adjoint representation. Therefore, we omit the subscript $adj$. We supplement  variables  $\cal T$ and $J$ with the upper \revisionZ{indexes $A,B$ that}  may take values $+$ or $-$. Then, for example, ${\cal T}^- = {\cal T}$ while ${\cal T}^+$ is its Hermitian conjugate. For the Hermitian conjugated Wilson lines we write
\begin{eqnarray}
{\cal T}^+_+(x^-,x_\bot) &=&\bar{P}\, {\rm exp}\,\Big(- i \int A^-(x^+,x^-,\vec{x}_\bot)d x^+ \Big)\nonumber\\
{\cal T}^+_-(x^+,x_\bot) &=& \bar{P}\, {\rm exp}\,\Big(- i \int A^+(x^+,x^-,\vec{x}_\bot)dx^- \Big)
\end{eqnarray}
Here on the right hand side the $\bar{P}$ notation means the reverse ordering of the integrals in the exponent's expansion . For brevity we denote below this reverse ordering by the same symbol $P$ as the direct one. \revisionU{More precisely, all expressions ${P}\, {\rm exp}\,\Big(-B i \int A^\pm dx^\mp \Big)$ with $B = +1$ should be understood as $\bar{P}\, {\rm exp}\,\Big(- i \int A^\pm dx^\mp \Big) $ while if $B=-1$ then we imply $P\, {\rm exp}\,\Big( i \int A^\pm dx^\mp \Big)$.} In the final expressions we will restore the notation $\bar{P}$.
\revisionU{We also introduce the following notations for some operators of interest:
\begin{eqnarray}
{\Phi}^-_+(x^-,x_\bot) &=&P\, {\rm exp}\,\Big( i \int_{-\infty}^\infty A^-(x^+,x^-,\vec{x}_\bot)d x^+ \Big) -1  \nonumber\\
{\Phi}^-_-(x^+,x_\bot) &=& P\, {\rm exp}\,\Big( i \int_{-\infty}^\infty A^+(x^+,x^-,\vec{x}_\bot)dx^- \Big)-1\nonumber
\end{eqnarray}
and
\begin{eqnarray}
{\Phi}^+_+(x^-,x_\bot) &=&\bar{P}\, {\rm exp}\,\Big(- i \int_{-\infty}^\infty A^-(x^+,x^-,\vec{x}_\bot)d x^+ \Big) -1 \nonumber\\
{\Phi}^+_-(x^+,x_\bot) &=& \bar{P}\, {\rm exp}\,\Big( -i \int_{-\infty}^\infty A^+(x^+,x^-,\vec{x}_\bot)dx^- \Big)-1\nonumber
\end{eqnarray}}

The partition function may be represented as follows
\begin{eqnarray}
Z  &=&\, \int DA \, {\rm exp}\,\Big(i S^0[A]
   \revisionU{+}{i\zeta_{AB}}{} \int d^2 x_\bot  \nonumber\\&&{\rm Tr}\,\Le P\,e^{-A\imath \,\int_{-\infty}^{\infty}\,d x^+\, \,A^-(x^+,0,x_\bot)\,}\,-\,1\,\Ra\, \partial_\bot^2\nonumber\\&&
\,\Le P\,e^{-B\imath \,\int_{-\infty}^{\infty}\,d x^-\, \,A^+(0,x^-,x_\bot)\,}\,-\,1\,\Ra\, \Big]\Big) \nonumber\\ &=& {\rm const}\, \int DA D{\cal A} \, {\rm exp}\,\Big(i S^0[A]  \,\nonumber\\&&
   \revisionU{-}i {\zeta}_{AB} \int d^4 x {\rm Tr}\, \Big[{{\cal A}^A_+}{} - {\delta(x^+)}{} \,\Le {\cal T}^A_+(x^-,x_\bot) \,-\,1\,\Ra\, \Big] \partial_\bot^2\nonumber\\&&
 \Big[  {{\cal A}^B_-(x)}{}-  \Le {\cal T}^B_-(x^+,x_\bot)-\,1\,\Ra\,{\delta(x^-)}{} \Big]\nonumber\\ &&
  \revisionU{+}{i\zeta_{AB}}{} \int d^2 x_\bot  {\rm Tr}\,\Le P\,e^{-A\imath \,\int_{-\infty}^{\infty}\,d x^+\, \,A^-(x^+,0,x_\bot)\,}\,-\,1\,\Ra\, \partial_\bot^2\nonumber\\&&
\,\Le P\,e^{-B\imath \,\int_{-\infty}^{\infty}\,d x^-\, \,A^+(0,x^-,x_\bot)\,}\,-\,1\,\Ra\, \Big]\Big)
 \nonumber\\&=&{\rm const}\, \int DA D{\cal A}\, {\rm exp}\,\Big(i S_{eff}[A,{\cal A}]\Big)\label{Z1}
\end{eqnarray}
Here $S_{eff}$ is an effective action for the interaction between the reggeons and the virtual gluons:
\begin{eqnarray}
&& S_{eff}[A,{\cal A}]  = S^{0}[A]
\revisionU{+}{\zeta_{AB}} {\rm Tr}\,\int d^4 x \partial_\bot {\cal A}_+^A\,\partial_\bot {\cal A}_-^B\nonumber\\&&
   \revisionU{+}\zeta_{AB}\int d^4 x \delta(x^+) {\rm Tr}\,\Phi_+^A \partial_\bot^2 {\cal A}_-^B(x)\nonumber\\&&
 \revisionU{+}\zeta_{AB}\int d^4 x \delta(x^-) {\rm Tr}\,\Phi_-^B \partial_\bot^2 {\cal A}_+^A(x)
\end{eqnarray}
\revisionU{which can be also represented as follows:}
\begin{eqnarray}
&&S_{eff}[A,{\cal A}]  = S^{0}[A]
\revisionU{+}{\zeta_{AB}} {\rm Tr}\,\int d^4 x \partial_\bot {\cal A}_+^A(x^+,x^-,x_\bot)\,\nonumber\\&&\partial_\bot {\cal A}_-^B(x^+,x^-,x_\bot)\label{SL}\\&&
  \revisionU{+}\frac{\zeta_{AB}}{2}{\rm Tr}\, \int d^4 x \partial_{x^+} \Bigl( P\,e^{-A\imath \,\int_{-\infty}^{x^+}\,d \bar{x}^+\, \,A^-(\bar{x}^+,x^-,x_\bot)\,}\nonumber\\&&-P\,e^{-A\imath \,\int^{\infty}_{x^+}\,d \bar{x}^+\, \,A^-(\bar{x}^+,x^-,x_\bot)\,}\,\Bigr)\, \partial_\bot^2 {\cal A}_-^B(0,x^-,\vec{x}_\bot)\nonumber\\&&
  \revisionU{+}\frac{\zeta_{AB}}{2}{\rm Tr}\, \int d^4 x \partial_{x^-} \,\Bigl( P\,e^{-B\imath \,\int_{-\infty}^{x^-}\,d \bar{x}^-\, \,A^+(x^+,\bar{x}^-, x_\bot)}\nonumber\\&&-P\,e^{-B\imath \,\int^{\infty}_{x^-}\,d \bar{x}^-\, \,A^+(x^+,\bar{x}^-, x_\bot)\,}\,\Bigr)\,  \partial_\bot^2 {\cal A}_+^A(x^+,0,\vec{x}_\bot)\nonumber
\end{eqnarray}
The fields coupled to the Wilson lines
$$\tilde{\cal A}^+(x^-,\vec{x}_\bot) = {\cal A}^+(0,x^-,\vec{x}_\bot),  \tilde{\cal A}^-(x^+,\vec{x}_\bot) = {\cal A}^-(x^+,0,\vec{x}_\bot)$$
satisfy the constraints
$$
\partial_\pm \tilde{\cal A}^\pm = \partial_\pm \tilde{\cal A}_\mp =0
$$
\revision{which guarantee, that $\tilde{\cal A}^\pm$ does not depend on $x^\pm$. At the same time the kinetic term in Eq. (\ref{SL}) for the reggeon field contains $\cal A$ rather than $\tilde{\cal A}$
We obtain that in the kinetic term  the field ${\cal A}^+(x^+,x^-,x_\bot)$ depends on $x^+$ while ${\cal A}^-(x^+,x^-,x_\bot)$ depends on $x^-$. It is worth mentioning, that the form of the action presented in \cite{Lipatov} contains the kinetic term, in which the reggeon fields depend only on three coordinates: the "$-$" component depends on $x^+,x_\bot$ while the "$+$" component depends on $x^-,x_\bot$. The consequences of this difference} \revisionZ{are analyzed in Appendix A, see also \eq{AddEq1}.}

\section{Effective action in the form of Lipatov}
\label{Lipatov2}

In terms of the constants $\zeta, \zeta^\prime$ the above derived action has the form (here in order to avoid the misunderstanding we denote the inverse ordering by $\bar{P}$)
\begin{eqnarray}
&& S_{eff}[A,{\cal A}]  = S^{0}[A]
 +{\rm Tr}\,\int d^4 x  \Big(\zeta \partial_\bot{\cal A}_+\,\partial_\bot {\cal A}_- \nonumber\\&& + \zeta \partial_\bot{\cal A}_+^+\,\partial_\bot {\cal A}_-^+ - \zeta^\prime \partial_\bot{\cal A}_+\,\partial_\bot {\cal A}_-^+-\zeta^\prime \partial_\bot{\cal A}_+^+\,\partial_\bot {\cal A}_- \Big)\nonumber\\&&
   \revisionU{+}\zeta\int d^4 x \delta(x^+) {\rm Tr}\,\Phi_+^-\, \partial_\bot^2 {\cal A}_-(x)\nonumber\\&&
 \revisionU{+}\zeta\int d^4 x \delta(x^-) {\rm Tr}\,\Phi_-^- \partial_\bot^2 {\cal A}_+(x)\nonumber\\&&
   \revisionU{-}\zeta^\prime\int d^4 x \delta(x^+) {\rm Tr}\,\Phi_+^+ \partial_\bot^2 {\cal A}_-(x)\nonumber\\&&
 \revisionU{-}\zeta^\prime\int d^4 x \delta(x^-) {\rm Tr}\,\Phi_-^+ \partial_\bot^2 {\cal A}_+(x)\nonumber\\&&
   \revisionU{+}\zeta\int d^4 x \delta(x^+) {\rm Tr}\,\Phi_+^+\partial_\bot^2 {\cal A}^+_-(x)\nonumber\\&&
 \revisionU{+}\zeta\int d^4 x \delta(x^-) {\rm Tr}\,\Phi_-^+\partial_\bot^2 {\cal A}^+_+(x)\nonumber\\&&
   \revisionU{-}\zeta^\prime\int d^4 x \delta(x^+) {\rm Tr}\,\Phi_+^- \partial_\bot^2 {\cal A}^+_-(x)\nonumber\\&&
 \revisionU{-}\zeta^\prime\int d^4 x \delta(x^-) {\rm Tr}\,\Phi_-^- \partial_\bot^2 {\cal A}^+_+(x)\label{evenmore}
\end{eqnarray}
For $\zeta$, $\zeta^\prime$ given by Eq. (\ref{naive00}) we obtain:
\begin{eqnarray}
&&S_{eff}[A,{\cal A}]  = S^{0}[A]
 \revisionU{+}\frac{1}{2g^2}{\rm Tr}\,\int d^4 x  \Big(\partial_\bot{\cal A}_+\,\partial_\bot {\cal A}_- \nonumber\\&& + \partial_\bot{\cal A}_+^+\,\partial_\bot {\cal A}_-^+ - \partial_\bot{\cal A}_+\,\partial_\bot {\cal A}_-^+-\partial_\bot{\cal A}_+^+\,\partial_\bot {\cal A}_- \Big)\nonumber\\&&
 \revisionU{+}\frac{\Delta \zeta}{2}{\rm Tr}\,\int d^4 x \Big(\partial_\bot{\cal A}_+\,\partial_\bot {\cal A}_- + \partial_\bot{\cal A}_+^+\,\partial_\bot {\cal A}_-^+\nonumber\\&& + \partial_\bot{\cal A}_+\,\partial_\bot {\cal A}_-^++ \partial_\bot{\cal A}_+^+\,\partial_\bot {\cal A}_- \Big)\label{more}\\&&
   \revisionU{+}\frac{1}{2g^2}\int d^4 x \delta(x^+) {\rm Tr}\,\Bigl( \Phi_+^- -\Phi_+^+\Bigr)\, \partial_\bot^2  ({\cal A}-{\cal A}^+) \nonumber\\&&
 \revisionU{+}\frac{1}{2g^2}\int d^4 x \delta(x^-) {\rm Tr}\,\Bigl( \Phi_-^- -\Phi_-^+\Bigr)\,  \partial_\bot^2 ({\cal A}-{\cal A}^+) \nonumber\\&&
   \revisionU{+}\frac{\Delta\zeta}{2}\int d^4 x \delta(x^+) {\rm Tr}\,\Bigl( \Phi_+^- +\Phi_+^+\Bigr)\, \partial_\bot^2 ({\cal A}_-+{\cal A}_-^+)\nonumber\\&&
 \revisionU{+}\frac{\Delta\zeta}{2}\int d^4 x \delta(x^-) {\rm Tr}\,\Bigl( \Phi_-^- +\Phi_-^+\Bigr)\,  \partial_\bot^2 ({\cal A}_++{\cal A}_+^+)\nonumber
\end{eqnarray}

The Hermitian version of the effective action in the form of Lipatov  \cite{Lipatov} corresponds to the simplest choice
$$
\zeta = \zeta^\prime = 1/(2g^2)
$$
Let us denote the anti - Hermitian part of $\cal A$ by $i {\cal B} = \frac{{\cal A}-{\cal A}^+}{2} = i B^a \lambda^a$. Then this choice of parameters gives
\begin{eqnarray}
&& S_{eff}[A,{\cal B}]  = S^{0}[A]
  \revisionU{-}\frac{2}{g^2}{\rm Tr}\,\int d^4 x \partial_\bot {\cal B}_+(x^+,x^-,x_\bot)\,\nonumber\\&&\partial_\bot {\cal B}_-(x^+,x^-,x_\bot) \\&&
   \revisionU{+}\frac{i}{g^2}\int d^4 x \delta(x^+) {\rm Tr}\,\Bigl( P\,e^{\imath \,\int_{-\infty}^{\infty}\,d x^+\, \,A^-(x^+,x^-,x_\bot)\,}\,\nonumber\\&&-\,\bar{P}\,e^{-\imath \,\int_{-\infty}^{\infty}\,d x^+\, \,A^-(x^+,x^-,x_\bot)\,}\Bigr)\, \partial_\bot^2  {\cal B}_-(x)\nonumber\\&&
 \revisionU{+}\frac{i}{g^2}\int d^4 x \delta(x^-) {\rm Tr}\,\Bigl( P\,e^{\imath \,\int_{-\infty}^{\infty}\,d x^-\, \,A^+(x^+,x^-,x_\bot)\,}\,\nonumber\\&&-\,\bar{P}\,e^{-\imath \,\int_{-\infty}^{\infty}\,d x^-\, \,A^+(x^+,x^-,x_\bot)\,} \,\Bigr)\,  \partial_\bot^2 {\cal B}_+(x)\nonumber
\end{eqnarray}
We should identify $\frac{1}{g}{\cal B}$ with the Hermitian $u(3)$ reggeon field.
The same expression can be written also in the more conventional form
\beqar\label{LipCur}
&&S_{eff}[A,{\cal B}]  =  S^{0}[A]
  \revisionU{-}\frac{2}{g^2}{\rm Tr}\,\int d^4 x \partial_\bot {\cal B}_+(x^+,x^-,x_\bot)\,\nonumber\\&&\partial_\bot {\cal B}_-(x^+,x^-,x_\bot)\nonumber \\
 &&\revisionU{+}
\frac{i}{2\,g^2}\int d^4 x\, \partial_{+}\, {\rm Tr}\,\Bigl(
P\,e^{\imath \,\int_{-\infty}^{x^{+}}\,d \bar{x}^+\, \,A^-(\bar{x}^+,x^-,x_\bot)\,}\,\nonumber\\&&-
P\,e^{\imath \,\int_{x^{+}}^{\infty}\,d x^+\, \,A^-(x^+,x^-,x_\bot)\,}\,\nonumber\\
&&-\bar{P}\,e^{-\imath \,\int_{-\infty}^{x^{+}}\,d \bar{x}^+\, \,A^-(\bar{x}^+,x^-,x_\bot)\,}\,\nonumber\\&&+
\bar{P}\,e^{-\imath \,\int_{x^{+}}^{\infty}\,d \bar{x}^+\, \,A^-(\bar{x}^+,x^-,x_\bot)\,}\,
\Bigr)\, \partial_\bot^2  {\cal B}_-(0,x^-,x_\bot)-\nonumber\\
&&\revisionU{+}
\frac{i}{2\,g^2}\int d^4 x \,\partial_{-}\, {\rm Tr}\,\Bigl(
P\,e^{\imath \,\int_{-\infty}^{x^{-}}\,d \bar{x}^-\, \,A^+(x^+,\bar{x}^-,x_\bot)\,}\,\nonumber\\&&-\,
P\,e^{\imath \,\int_{x^{-}}^{\infty}\,d \bar{x}^-\, \,A^+(x^+,\bar{x}^-,x_\bot)\,} \nonumber \\
&&-\,\bar{P}\,e^{-\imath \,\int_{-\infty}^{x^{-}}\,d \bar{x}^-\, \,A^+(x^+,\bar{x}^-,x_\bot)\,}\,\nonumber\\&&+\,
\bar{P}\,e^{-\imath \,\int_{x^{-}}^{\infty}\,d \bar{x}^-\, \,A^+(x^+,\bar{x}^-,x_\bot)\,} \,
\Bigr)\,  \partial_\bot^2 {\cal B}_+(x^+,0,x_\bot)
\eeqar
\revision{Equivalently, we may rewrite it as
\beqar
&& S_{eff}[A,{\cal B}]  =  S^{0}[A]
  \revisionU{-}\frac{2}{g^2}{\rm Tr}\,\int d^4 x \partial_\bot {\cal B}_+(x^+,x^-,x_\bot)\,\nonumber\\&&\partial_\bot {\cal B}_-(x^+,x^-,x_\bot) \label{Lipat}\\
 &&\revisionU{+}
\frac{i}{\,g^2}\int d^4 x\, \partial_{+}\, {\rm Tr}\,\Bigl(
P\,e^{\imath \,\int_{-\infty}^{x^{+}}\,d \bar{x}^+\, \,A^-(\bar{x}^+,x^-,x_\bot)\,}\,\nonumber\\&&-
\bar{P}\,e^{-\imath \,\int_{-\infty}^{x^{+}}\,d \bar{x}^+\, \,A^-(\bar{x}^+,x^-,x_\bot)\,}\,
\Bigr)\, \partial_\bot^2  {\cal B}_-(0,x^-,x_\bot)\nonumber\\
&&\revisionU{+}
\frac{i}{\,g^2}\int d^4 x \,\partial_{-}\, {\rm Tr}\,\Bigl(
P\,e^{\imath \,\int_{-\infty}^{x^{-}}\,d \bar{x}^-\, \,A^+(x^+,\bar{x}^-,x_\bot)\,}\, \nonumber\\&&-
\,\bar{P}\,e^{-\imath \,\int_{-\infty}^{x^{-}}\,d \bar{x}^-\, \,A^+(x^+,\bar{x}^-,x_\bot)\,}\,
\Bigr)\,  \partial_\bot^2 {\cal B}_+(x^+,0,x_\bot)\nonumber
\eeqar}
\revision{ It was mentioned above, that the dependence of ${\cal B}^\pm$ on $x^\pm$ along the light cone is suppressed at} \revisionZ{ $|s/t| \gg 1$. We, therefore, represent
the reggeon field} as
\beq\label{AddEq1}
{\cal B}^+(x^+,x^-,x_\bot) = {\cal B}^+(0,x^-,x_\bot) + {\cal D}^+(x^+,x^-,x_\bot)\eeq \beq {\cal B}^-(x^+,x^-,x_\bot) = {\cal B}^-(x^+,0,x_\bot) + {\cal D}^-(x^+,x^-,x_\bot)
\eeq
where the \revisionZ{typical deviation of $x^{\pm}$ from zero in ${\cal B}^\pm$ may be estimated as
\beq\label{AddEq11}
x^{\pm}\,\propto\,\frac{1}{\sqrt{s}}
\eeq}
The fields ${\cal D}^\pm$ obey the constraint
$$
{\cal D}^+(0,x^-,x_\bot)={\cal D}^-(x^+,0,x_\bot)=0
$$
and the redefined action receives the following form:
\beqar
&& S_{eff}[A,{\cal B}]  =  S^{0}[A]
  \revisionU{-}\frac{2}{g^2}{\rm Tr}\,\int d^4 x \partial_\bot {\cal B}_+(x^+,0,x_\bot)\,\nonumber\\&&\partial_\bot {\cal B}_-(0,x^-,x_\bot)\label{Lipat1}\\ && \revisionU{-}  \frac{2}{g^2}{\rm Tr}\,\int d^4 x \partial_\bot {\cal D}_+(x^+,x^-,x_\bot)\,\partial_\bot {\cal D}_-(x^+,x^-,x_\bot)\nonumber\\ && \revisionU{-} \frac{2}{g^2}{\rm Tr}\,\int d^4 x \partial_\bot {\cal D}_+(x^+,x^-,x_\bot)\,\partial_\bot {\cal B}_-(0,x^-,x_\bot)\nonumber\\&&\revisionU{-} \frac{2}{g^2}{\rm Tr}\,\int d^4 x \partial_\bot {\cal B}_+(x^+,0,x_\bot)\,\partial_\bot {\cal D}_-(x^+,x^-,x_\bot)\nonumber\\
 &&\revisionU{+}
\frac{i}{\,g^2}\int d^4 x\, \partial_{+}\, {\rm Tr}\,\Bigl(
P\,e^{\imath \,\int_{-\infty}^{x^{+}}\,d \bar{x}^+\, \,A^-(\bar{x}^+,x^-,x_\bot)\,}\,\nonumber\\&&-
\bar{P}\,e^{-\imath \,\int_{-\infty}^{x^{+}}\,d \bar{x}^+\, \,A^-(\bar{x}^+,x^-,x_\bot)\,}\,
\Bigr)\, \partial_\bot^2  {\cal B}_-(0,x^-,x_\bot)\nonumber\\
&&\revisionU{+}
\frac{i}{\,g^2}\int d^4 x \,\partial_{-}\, {\rm Tr}\,\Bigl(
P\,e^{\imath \,\int_{-\infty}^{x^{-}}\,d \bar{x}^-\, \,A^+(x^+,\bar{x}^-,x_\bot)\,}\, \nonumber\\&&-
\,\bar{P}\,e^{-\imath \,\int_{-\infty}^{x^{-}}\,d \bar{x}^-\, \,A^+(x^+,\bar{x}^-,x_\bot)\,}\,
\Bigr)\,  \partial_\bot^2 {\cal B}_+(x^+,0,x_\bot)\nonumber
\eeqar
One can see, that the field $\cal D$ does not interact directly with the gluon field $A$.
\revisionZ{On the diagrammatic levels the emission and absorbtion of the quanta of $\cal D$ by $\cal B$ modifies the propagator of the latter field (see Appendix A). In Appendix B we present the form of the theory, which coincides with the present one in the leading orders but, in which the field $\cal D$ interact directly with virtual gluons.}

In the kinetic term for the reggeon field the coefficient of Eq. (210) of \cite{Lipatov} is twice larger than that of our Eq. (\ref{Lipat}) or of Eq. (7) of \cite{Lipatoveff2}. However, in Eq. (221) of \cite{Lipatov} the necessity to perform the finite renormalization has been pointed out. The resulting kinetic term corresponds to our expression of Eq. (\ref{Lipat}). The alternative explanation of this finite renormalization is given, for example, in \cite{Bondarenko:2017vfc}. In this approach the initial kinetic coefficient of Eq. (210) in \cite{Lipatov} may be kept while the perturbation expansion around the classical solution $A = {\cal B}$ gives as a result our coefficient in front of the kinetic term in Eq. (\ref{Lipat}). In the leading order this corresponds to our prescription, which allows to avoid the overcounting in the perturbation theory. The reggeon field itself is in essence a composition of the gluons corresponding to certain classical configuration of the field. It would be counted twice if we construct the perturbation theory around the classical solution $A = {\cal B}$.
\revisionZ{ Thus in order to avoid the overcounting we simply construct the perturbation theory around $A = 0$. Equivalently, we may use the following modification of the action, in which the coefficient in front of the kinetic term of the reggeon field is doubled:
\beqar
&\,&S_{eff}[A,{\cal B}] =  S^{0}[A]
  \revisionU{-}\frac{4}{g^2}{\rm Tr}\,\int d^4 x \partial_\bot {\cal B}_+(x^+,x^-,x_\bot)\,\nonumber\\&&\partial_\bot {\cal B}_-(x^+,x^-,x_\bot) \label{Lipat2}\\
 &&\revisionU{+}
\frac{i}{\,g^2}\int d^4 x\, \partial_{+}\, {\rm Tr}\,\Bigl(
P\,e^{\imath \,\int_{-\infty}^{x^{+}}\,d \bar{x}^+\, \,A^-(\bar{x}^+,x^-,x_\bot)\,}\,\nonumber\\&&-
\bar{P}\,e^{-\imath \,\int_{-\infty}^{x^{+}}\,d \bar{x}^+\, \,A^-(\bar{x}^+,x^-,x_\bot)\,}\,
\Bigr)\, \partial_\bot^2  {\cal B}_-(0,x^-,x_\bot)\nonumber\\
&&\revisionU{+}
\frac{i}{\,g^2}\int d^4 x \,\partial_{-}\, {\rm Tr}\,\Bigl(
P\,e^{\imath \,\int_{-\infty}^{x^{-}}\,d \bar{x}^-\, \,A^+(x^+,\bar{x}^-,x_\bot)\,}\, \,\nonumber\\&&-
\,\bar{P}\,e^{-\imath \,\int_{-\infty}^{x^{-}}\,d \bar{x}^-\, \,A^+(x^+,\bar{x}^-,x_\bot)\,}\,
\Bigr)\,  \partial_\bot^2 {\cal B}_+(x^+,0,x_\bot)\nonumber
\eeqar}
\revisionZ{Working with this action we construct the perturbation theory around the classical solution $A = {\cal B}$ (obtained in the leading order of the expansion of the Wilson lines in powers of $A$). The action of Eq. (\ref{Lipat2}) corresponds to the Hermitian version of Eq. (210) of \cite{Lipatov}. At the same time Eq. (\ref{Lipat}) corresponds to the form of the action used in \cite{Lipatoveff2} (see also  \cite{Antonov:2004hh}). Notice, that the Hermitian form of Lipatov action has been used also in \cite{Nefedov}. We may also write this action in the following form:
\beqar\label{Lipat4}
&& S_{eff}[A,{\cal B}]  =  S^{0}[A]
  \revisionU{-}\frac{4}{g^2}{\rm Tr}\,\int d^4 x \partial_\bot {\cal B}_+(x^+,x^-,x_\bot)\,\nonumber\\&&\partial_\bot {\cal B}_-(x^+,x^-,x_\bot)\nonumber \\
 &&\revisionU{+}
\frac{i}{2\,g^2}\int d^4 x\, \partial_{+}\, {\rm Tr}\,\Bigl(
P\,e^{\imath \,\int_{-\infty}^{x^{+}}\,d \bar{x}^+\, \,A^-(\bar{x}^+,x^-,x_\bot)\,}\,\nonumber\\&&-
P\,e^{\imath \,\int_{x^{+}}^{\infty}\,d x^+\, \,A^-(x^+,x^-,x_\bot)\,}\nonumber\\
&&-
\bar{P}\,e^{-\imath \,\int_{-\infty}^{x^{+}}\,d \bar{x}^+\, \,A^-(\bar{x}^+,x^-,x_\bot)\,}\,\nonumber\\&&+
\bar{P}\,e^{-\imath \,\int_{x^{+}}^{\infty}\,d \bar{x}^+\, \,A^-(\bar{x}^+,x^-,x_\bot)\,}\,
\Bigr)\, \partial_\bot^2  {\cal B}_-(0,x^-,x_\bot)-\nonumber\\
&&\revisionU{+}
\frac{i}{2\,g^2}\int d^4 x \,\partial_{-}\, {\rm Tr}\,\Bigl(
P\,e^{\imath \,\int_{-\infty}^{x^{-}}\,d \bar{x}^-\, \,A^+(x^+,\bar{x}^-,x_\bot)\,}\,\nonumber\\&&-\,
P\,e^{\imath \,\int_{x^{-}}^{\infty}\,d \bar{x}^-\, \,A^+(x^+,\bar{x}^-,x_\bot)\,} \nonumber \\
&&-
\,\bar{P}\,e^{-\imath \,\int_{-\infty}^{x^{-}}\,d \bar{x}^-\, \,A^+(x^+,\bar{x}^-,x_\bot)\,}\,\nonumber\\&&+\,
\bar{P}\,e^{-\imath \,\int_{x^{-}}^{\infty}\,d \bar{x}^-\, \,A^+(x^+,\bar{x}^-,x_\bot)\,} \,
\Bigr)\,  \partial_\bot^2 {\cal B}_+(x^+,0,x_\bot)
\eeqar}

\revision{In order to use the above actions for the calculation of the leading contribution to the scattering amplitude} of the $i$ - the sector in the multi - Regge kinematics (with the $i-1$ - the and $i$ - th produced high energy particles) we should calculate the correlator of Eq. (\ref{Dq2}) in the center of intertia reference frame of the two colliding hadrons. Suppose, that in this reference frame the worldlines of the created particles are along the vectors $\vec{n}$ and $\vec{n}^\prime$. Then we have
\begin{eqnarray}
{\cal D}(-q^2)&=&\frac{1}{ Z^\prime }\int DA D{\cal B} e^{i S_{eff}[A,{\cal B}]} \,
\int \,d^{2}\vec{x}_\bot\,e^{-\imath\,\Le\,{q}{x}_\bot\Ra}\, \nonumber\\&&\Le\,P\,e^{\imath \,\int_{-\infty}^{\infty}\,d \lambda\, n_\mu A^\mu(x_\bot + \lambda n)\,}\,-\,1\,\Ra\, \,\nonumber\\&&\otimes
\,\Le\,P\,e^{\imath \,\int_{-\infty}^{\infty}\,d \lambda\, n^\prime_\mu A^\mu(\lambda n^\prime) \,}\,-\,1\,\Ra\,\label{Dq22}
\end{eqnarray}
where  $x_\bot$ belongs to the plane given by the conditions $nx_\bot=n^\prime x_\bot =0$.
Following \cite{Lipatov} we assume that in Eq. (\ref{Dq2}) we should restrict ourselves by the excitations from the small but finite region of rapidities (to which the momenta of the two considered produced particles belong). However, one can see, that here there is no simple expression for $D(-q^2)$ in terms of the propagator of the reggeon field. In the next section we will see, that such a representation does exist if we consider the $i$ - sector of multi - Regge kinematics alone neglecting its correlation with the other produced particles as well as with the colliding hadrons.

\section{The effective theory for the $i$ - th sector of the multi - Regge kinematics}
\label{sector}

\revisionF{For the multi - Regge kinematics the momenta of the two initial partons are $p_A$ and $p_B$. The final state gluon momenta are $k_0 = p_A^\prime , k_1, ..., k_n, k_{n+1} = p_B^\prime$. we have
\begin{equation}
s \gg  s_i = 2k_{i-1}k_i \gg t_i = q^2_i = \Big(p_A - \sum_{r=0}^{i-1}k_r\Big)^2\nonumber
\end{equation}
And in the leading logarithmic approximation the multi - gluon production amplitude has the form:
\begin{equation}
A_{LLA} \sim \Pi_{i=1}^{n+1} s_i^{\omega(t_i)}\label{0r2}
\end{equation}
Here $\omega(t)$ is given by Eq. (\ref{1r}). }

\revisionF{The rapidities of the physical gluons in the above sequence are ordered according to the ordering of momenta $k_i$. Let us consider what is happening to the pair of the adjacent physical gluons in the above sequence (i.e. those, which give rise to jets and split into the partons of the final hadron states and have the momenta $k_i$ and $k_{i-1}$). Using the fact that $k_i$ and $k_{i-1}$ are not collinear}, we may come to the reference frame, where the total $3$ momentum of the pair is zero. In this reference frame the physical gluons are moving in the opposite directions. We choose the coordinates of this reference frame in such a way, that the $x$ axis is directed along the momentum of the first of the two considered gluons. The corresponding vector $n$ is $(1,1,0,0)$. The kinematics implies that the momentum transfer $q$ has the components $q^0=0$ and $\vec{q} = \vec{q}_\bot \bot \vec{n}$. We may chose the reference frame, in which $\vec{n} = (1,0,0)$. According to the common lore \revisionF{(see \cite{8,9,10,4,13,14,Balitsky:1978ic})} the $i$ - th sector of the Regge kinematics may be considered separately from the other produced particles and without any relation to the initial colliding hadrons.

Then we may construct the effective 2D model for the Wilson lines taken along the light cone coordinates corresponding to vector $\vec{n}$, which marks the motion of the $i-1$ - th and the $i$ - th produced particles in their center of inertia reference frame. The contribution to the overall amplitude from the interaction of the $i$-th and the $i-1$ - th particles is given by the function $D(q_\bot)$ of Eq. (\ref{Dq2}) with the action supplemented by the term of Eq. (\ref{S+-2}). The Fourier transform of this function may be written as follows
\begin{eqnarray}
&&{\bf D}^{(i,i-1)}_{ab}(x_\bot)  = \frac{1}{Z^\prime} \int DA \, {\rm exp}\,\Big(i S^{0}[A]\nonumber\\&&
  \revisionU{-}\sum_{A,B=\pm}{i\zeta_{AB}}{} \int d^2 x_\bot  {\rm Tr}\,\partial_\bot\Le P\,e^{-A\imath \,\int_{-\infty}^{\infty}\,d x^+\, \,A^-(x^+,0,\vec{x}_\bot)\,}\,\Ra\,\nonumber\\&& \partial_\bot\Le P\,e^{-B\imath \,\int_{-\infty}^{\infty}\,d x^-\, \,A^+(0,x^-,\vec{x}_\bot)\,}\,\,\Ra\, \Big)\nonumber\\&&
\Bigl({\rm Tr}\,\Big[\lambda^a\,\Le P\,e^{\imath \,\int_{-\infty}^{\infty}\,d x^+\, \,A^-(x^+,0,x_\bot)\,}\,-\,1\,\Ra\,\Big]\nonumber\\&& {\rm Tr}\,\Big[\lambda^b\,
\,\Le P\,e^{\imath \,\int_{-\infty}^{\infty}\,d x^-\, \,A^+(0,x^-,0)\,}\,-\,1 \Ra\, \Big]\Bigr)\nonumber
\end{eqnarray}
\revisionF{In the following instead of this correlator we will use its symmetric modification
\begin{eqnarray}
&&{D}^{(i,i-1)}_{ab}(x_\bot)  = \frac{1}{\revisionFF{4} Z^\prime} \int DA \, {\rm exp}\,\Big(i S^{0}[A]\nonumber\\&&
  \revisionU{-}\sum_{A,B=\pm}{i\zeta_{AB}}{} \int d^2 x_\bot  {\rm Tr}\,\partial_\bot\Le P\,e^{-A\imath \,\int_{-\infty}^{\infty}\,d x^+\, \,A^-(x^+,0,\vec{x}_\bot)\,}\,\Ra\,\nonumber\\&& \partial_\bot\Le P\,e^{-B\imath \,\int_{-\infty}^{\infty}\,d x^-\, \,A^+(0,x^-,\vec{x}_\bot)\,}\,\,\Ra\, \Big)\nonumber\\&&
\Bigl({\rm Tr}\,\Big[\lambda^a\,\Bigl( P\,e^{\imath \,\int_{-\infty}^{\infty}\,d x^+\, \,A^-(x^+,0,x_\bot)\,}\,\nonumber\\&&-\,\bar{P}\,e^{-\imath \,\int_{-\infty}^{\infty}\,d x^+\, \,A^-(x^+,0,x_\bot)\,}\,\Bigr)\,\Big]\nonumber\\&& {\rm Tr}\,\Big[\lambda^b\,
\,\Bigl( P\,e^{\imath \,\int_{-\infty}^{\infty}\,d x^-\, \,A^+(0,x^-,0)\,}\,\nonumber\\&&-\,\bar{P}\,e^{-\imath \,\int_{-\infty}^{\infty}\,d x^-\, \,A^+(0,x^-,0)\,} \Bigr)\, \Big]\Bigr)\label{D3}
\end{eqnarray}
Due to the time reversal symmetry the physical amplitude for the processes with the multi - Regge kinematics may be expressed through the correlator of this form. As in Sect. \ref{Lipatov2} the Hermitian version of the effective action in the form of Lipatov  \cite{Lipatov} corresponds to the simplest choice
$$
\zeta = \zeta^\prime = 1/(2g^2)
$$
Below we assume this choice of parameters,} \revisionFF{which means that
$$\zeta_{AB} = \left(\begin{array}{cc}\zeta & -\zeta \\ -\zeta & \zeta \end{array}\right)$$
At the same time in the majority of expressions below we keep tensor $\zeta_{AB}$ and do not express it explicitly through scalar $\zeta$. This allows to represent those expressions in a compact way and reveals the similarity with the expressions derived in the previous sections.
}

Due to the isotropy of space - time the function $D$ cannot depend on the direction of $\vec{q}_\bot$, and we are able to write:
\begin{equation}
D^{(i,i-1)}(\vec{q}_\bot) = {\cal D}^{(i,i-1)}(\vec{q}^2) = {\cal D}^{(i,i-1)}(-q^2)
\end{equation}
where we use, that $q^0=0$ in the given reference frame. This expression allows to calculate $D$ in the original reference frame, where the momenta of the two scattered particles are not collinear.

Next, we  define the generating functional
\begin{eqnarray}
&& Z[J]  =\, \int DA \, {\rm exp}\,\Big(i S^0[A]
   \revisionU{+}{i\zeta_{AB}}{} \int d^2 x_\bot  {\rm Tr}\,\nonumber\\&&\Le P\,e^{-A\imath \,\int_{-\infty}^{\infty}\,d x^+\, \,A^-(x^+,0,x_\bot)\,}\,-\,1+J^A_+(x_\bot)\frac{1}{\partial_\bot^2}\,\Ra\, \partial_\bot^2\nonumber\\&&
\,\Le P\,e^{-B\imath \,\int_{-\infty}^{\infty}\,d x^-\, \,A^+(0,x^-,x_\bot)\,}\,-\,1 + \frac{1}{\partial_\bot^2} J^B_-(x_\bot)\,\Ra\, \Big]\nonumber\\&&\revisionU{-} {i\zeta_{AB}}{} \int d^2 x_\bot  {\rm Tr}\,J^A_+(x_\bot)\frac{1}{\partial_\bot^2} J^B_-(x_\bot)\Big)\nonumber\\
 &=& {\rm const}\, \int DA D{\cal A} \, {\rm exp}\,\Big(i S^0[A] \nonumber\\&& \revisionU{-}{i \zeta_{AB}}{} \int d^2 x_\bot  {\rm Tr}\,J^A_+(x_\bot)\frac{1}{\partial_\bot^2} J^B_-(x_\bot)\,\nonumber\\&&
   \revisionU{-}i {\zeta}_{AB} \int d^4 x {\rm Tr}\, \nonumber\\&& \Big[{{\cal A}^A_+}{} - {\delta(x^+)}{} \,\Le {\cal T}^A_+(x^-,x_\bot) \,-\,1+J^A_+(x_\bot)\frac{1}{\partial_\bot^2}\,\Ra\, \Big] \partial_\bot^2\nonumber\\&&
 \Big[  {{\cal A}^B_-(x)}{}-  \Le {\cal T}^B_-(x^+,x_\bot)-\,1 + \frac{1}{\partial_\bot^2}J^B_-(x_\bot)\,\Ra\,{\delta(x^-)}{} \Big]\nonumber\\ &&
  \revisionU{+}{i\zeta_{AB}}{} \int d^2 x_\bot  {\rm Tr}\,\nonumber\\&&\Le P\,e^{-A\imath \,\int_{-\infty}^{\infty}\,d x^+\, \,A^-(x^+,0,x_\bot)\,}\,-\,1+J^A_+(x_\bot)\frac{1}{\partial_\bot^2}\,\Ra\, \partial_\bot^2\nonumber\\&&
\,\Le P\,e^{-B\imath \,\int_{-\infty}^{\infty}\,d x^-\, \,A^+(0,x^-,x_\bot)\,}\,-\,1 + \frac{1}{\partial_\bot^2} J^B_-(x_\bot)\,\Ra\, \Big]\Big)\nonumber
\end{eqnarray}
We have
\begin{eqnarray}\label{BarProp1}
&& D^{(i,i-1)}_{ab}(x_\bot-y_\bot)  =  -\, \frac{\delta^2}{\revisionF{4\zeta^2} \delta J^a_+(x_\bot) \delta J^a_-(y_\bot)} \,{\rm log}\, Z[J]\Big|_{J=0}\nonumber\\
&&= - \frac{\,\delta^2}{ 4\zeta^2\delta J^a_+(x_\bot) \delta J^b_-(y_\bot)}\,{\rm log}\,\int DA D{\cal A}\, \nonumber\\&&{\rm exp}\,\Big(i S^0[A]  \revisionU{-} {i \zeta_{AB}} \int d^2 x_\bot  {\rm Tr}\,J^A_+(x_\bot)\frac{1}{\partial_\bot^2} J^B_-(x_\bot)\,\nonumber\\&& \revisionU{+}{i\zeta_{AB}} {\rm Tr}\,\int d^4 x \partial_\bot {\cal A}_+^A\,\partial_\bot {\cal A}_-^B\nonumber\\&&
 \revisionU{+}i \zeta_{AB} \int d^4 x \delta(x^+) {\rm Tr}\, \Phi_+^A \partial_\bot^2 {\cal A}^B_-(x^+,x^-,x_\bot)\nonumber\\&&
\revisionU{+}i \zeta_{AB}\int d^4 x \delta(x^-) {\rm Tr}\,\Phi_-^B \partial_\bot^2 {\cal A}_+^A(x^+,x^-,x_\bot)
\nonumber\\&&
  \revisionU{+}i \zeta_{AB} \int d^4 x \delta(x^+) {\rm Tr}\,J^A_+(x_\bot) {\cal A}_-^B(x^+,x^-,x_\bot)\nonumber\\&&
  \revisionU{+}i \zeta_{AB}\int d^4 x \delta(x^-) {\rm Tr}\, J^B_-(x_\bot) {\cal A}_+^A(x^+,x^-,x_\bot)\Big)_{J=0}
 \nonumber\\
&&= \, \frac{1}{\revisionFF                                  {4\zeta^2}Z(0)}\, \int DA D{\cal A}\, {\rm exp}\,\Big(i S^0[A]\nonumber\\&&  \revisionU{+} {i\zeta_{AB}} {\rm Tr}\,\int d^4 x \partial_\bot {\cal A}_+^A\,\partial_\bot {\cal A}_-^B\nonumber\\&&
 \revisionU{+}i \zeta_{AB} \int d^4 x \delta(x^+) {\rm Tr}\,\Phi_+^A \partial_\bot^2 {\cal A}_-^B(x)\nonumber\\&&
 \revisionU{+}i \zeta_{AB} \int d^4 x \delta(x^-) {\rm Tr}\,\Phi_-^B \partial_\bot^2 {\cal A}_+^A(x)\Big)\nonumber\\&&
 \Big(\revisionF{\zeta^{-1,A}\zeta^{-1,B}\int dx^- {\cal A}_{-,a}^A(0,x^-,x_\bot)\int dy^+
  {\cal A}_{+,b}^B(y^+,0,y_\bot)}\nonumber\\&& \revisionU{+} \frac{i\zeta \delta_{ab}}{ \partial_\bot^2}\Big)
\end{eqnarray}
The second term in the brackets of the last row corresponds to bare propagator $1/q_\bot^2$, which is not essential, when we consider the multi - gluon exchange that should give the result enhanced by the factor $\Big(\frac{s}{t}\Big)^{\omega(q_\bot)}$. Therefore, this term may be omitted in our considerations.
We come to the conclusion, that  \revisionZ{ the Fourier transform of $D(x_\bot)$ may be considered as the propagator of the reggeon field ${\cal A}^\pm$ up to the bare term $\frac{i\,(2\pi)^2}{ q_\bot^2}$:
\begin{eqnarray}
&& \revisionFF{4\zeta^2}D^{(i,i-1)}_{ab}(q_\bot)  =  \revisionU{-}\frac{i\zeta\,(2\pi)^2}{\,q_\bot^2}\,\delta_{ab} \nonumber\\&&\revisionU{+}\frac{1}{Z(0)}\,\int d^2 x_\bot dy^+ dx^- e^{-i \vec{q}_\bot \vec{x}_\bot}\revisionF{\zeta^{-1,A}\zeta^{-1,B} } \nonumber\\&&\int DA D{\cal A}\, {\rm exp}\,\Big(i S_{eff}[A,{\cal A}]\Big)\revisionF{{\cal A}_{-,a}^A(0,x^-,x_\bot)
  {\cal A}_{+,b}^B(y^+,0,0)}\nonumber
\end{eqnarray}}

We denote the anti - Hermitian part of $\cal A$ by $i {\cal B} = \frac{{\cal A}-{\cal A}^+}{2} = i B^a \lambda^a$ and using Eq. (\ref{Lipat}) we rewrite the leading contribution of the $i-1$ -th and the $i$ -th produced particles to the scattering amplitude as follows
\begin{eqnarray}
&& D^{(i,i-1)}_{ab}(q_\bot) =  \revisionU{+}\frac{i\,(2\pi)^2}{\revisionFF{4\zeta}\,q_\bot^2}\,\delta_{ab} \label{DD}\\&&\revisionU{-} \frac{\revisionFF{1}}{Z(0)}\,\int d^2 x_\bot dy^+ dx^- e^{-i \vec{q}_\bot \vec{x}_\bot}  \nonumber\\&&\int DA D{\cal B}\, {\rm exp}\,\Big(i S_{eff}[A,{\cal B}]\Big){\cal B}_{-}^{a}(0,x^-,x_\bot)
  {\cal B}_{+}^{b}(y^+,0,0)\nonumber \label{DD}
\end{eqnarray}
\revisionZ{Recall, that using the action of the form of Eq. (\ref{Lipat}) or Eq. (\ref{Lipat1}) we should build the perturbation theory in such a way, that the field $A$ is expanded around $A=0$ although in the presence of the reggeon field $\cal B$ the classical solution is $A = {\cal B}$. On the other hand, we may use the action of the form of Eq. (\ref{Lipat2}). Then the perturbation theory is to be constructed around the leading order classical solution $A = {\cal B}$. The results of the calculations coincide identically \cite{Lipatov,Bondarenko:2017vfc}.}

\revisionZ{Notice, that for the action of the form close to that of Eq. (\ref{Lipat1}) the leading order reggeon propagator has been calculated in \cite{Bondarenko:2017vfc} (with the contributions of $\cal D$ neglected and with the forward Wilson line $\cal T$ instead of the anti - Hermitian combination $\frac{1}{2}({\cal T} - {\cal T}^+)$). The result of  \cite{Bondarenko:2017vfc} is
\begin{equation}
D^{(i,i-1)}(q_\bot) \sim s_i^{\omega(q_\bot)}
\end{equation}
for $\omega$ given by Eq. (\ref{1r}) with $t = q^2$. We checked that this asymptotic behavior is not changed if we substitute the anti - Hermitian combination $\frac{1}{2}({\cal T} - {\cal T}^+)$ as in Eq. (\ref{Lipat1}) (see \cite{Nefedov}, where also the Hermitian version of Lipatov action was considered). As for the field $\cal D$, it is also suppressed in the Regge kinematical region and may give the sub - leading corrections only. Moreover, the whole theory may be reformulated in such a way, that this field interacts with the gluons directly (see Appendix B). Again, with such a definition, the leading order asymptotic expressions for the scattering amplitudes in the multi - Regge kinematics should remain the same, but the sub - leading corrections may be different. }

\section{Conclusions}

\label{concl}

In this paper we discuss the high energy hadron - hadron scattering processes with the multi - Regge kinematics. The particles produced in such processes are mostly gluons, which give rise to jets and split (during the sequence of the other processes) to the partons of the created hadrons. We refer to those gluons as to the physical ones. During the scattering process the physical gluons with the close values of rapidities interact with each other via the exchange by multiple virtual gluons. In the literature this process is often referred to as the radiation of soft gluons by the hard (physical) gluons. It is worth mentioning that considering this process we may assume that the two interacting hard gluons were not created during the inelastic scattering, but existed forever and just scatter one on another. This does not change the asymptotic expression (for large $s_i/t_i$) for the contribution to the overall inelastic scattering amplitude.

The interaction between the created particles with close rapidities is the subject of an effective theory invented by Lipatov, in which the composite multi - gluon object called reggeon interacts with the conventional virtual gluons \cite{Lipatov}. The use of this theory allows to sum the essentially smaller number of diagrams in order to achieve the needed accuracy of calculations than the usual perturbation theory of QCD. Several forms of the effective action that are similar but not completely equivalent were proposed in different papers \cite{LipatovEff}. In the present paper we propose the relatively simple derivation of a new version of the effective action for this theory. The action, which is derived above, is reduced to the hermitian form of the action proposed in \cite{Lipatov} for a certain (and rather natural) choice of the effective coupling constants. However, for the other choices of coupling constants (to be obtained from the experiment) the action derived above differs somehow from the form of Lipatov. We suppose, that it may be considered as its improvement.

Moreover, we consider the two versions of the effective theory for the reggeons interacting with the gluons. The more traditional version is written in the light - cone coordinates corresponding to the direction of motion of the two colliding hadrons in their center of mass reference frame. This version directly corresponds to the Lipatov effective theory. \revision{However, within our version of this theory we are not able to write immediately the simple expression for the contribution of the interaction between the $i$ - th and the $i-1$ - th produced particles to the overall scattering amplitude in terms of the propagator of reggeon. In order to achieve this goal we propose the version of the theory that operates with the $i$ - th and the $i-1$ - th produced particles as if they were actually colliding instead of the actually colliding hadrons. The justification for this consideration is the factorization property of the amplitudes discussed in some details, for example, in  \cite{8,9,10,4,13,14}. Then we come to the reference frame, where the total 3 - momentum of the $i$ - th and the $i-1$ - th particles is zero. In this reference frame we define the light cone coordinates corresponding to the direction of motion of those two particles. In these light cone coordinates we come to the effective theory, in which the contribution of the $i$ - th and the $i-1$ - th produced physical gluons to the overall amplitude is expressed directly through the propagator of the reggeon.}

The essence of our derivation is the combination of the dimensional reduction $3+1$ D $\to 2$ D specific for the high energy scattering with the resummation of the corrections due to the virtual gluons in the Eikonal approximation. Considering the dimensional reduction we start from the approach of E. Verlinde, H. Verlinde \cite{Verlinde} and the approach of I.Aref'eva \cite{Arefeva}. However, instead of following their derivations, we propose our own simple way to deduce the effective $2D$ theory for the high energy scattering of the two high energy particles    (this is the scattering, in which the Mandelstam variables obey the relation $s \gg t$). Namely, in the reference frame, where the total $3$ momentum of the colliding particles is zero, they move mostly along the light cone $x^\pm$. This high energy motion assumes, that during the collision the gauge field $A$ does not have time to change. It may, therefore, be considered as constant in a certain approximation. In order to give the description of this approximation we are able to consider the gauge theory in $3+1$ D space with the values of light cone coordinates $x^\pm$ that belong to an interval $0<x^\pm <\Delta$, where $\Delta \sim \frac{1}{\sqrt{s}}$. The momentum transfer $\sqrt{-t}$ sets the distance corresponding to the typical variations of the gauge field $\delta \sim \frac{1}{\sqrt{t}} \gg \Delta$. In the zero approximation we disregard them. As a result we come to the action of the $2D$  model Eq. (\ref{S+-}) with the constants given by Eq. (\ref{naive00}). The value of $\Delta \zeta$ remains arbitrary even in this simplified consideration. The given  model operates with the Wilson lines that depend on the coordinates in the plane orthogonal to the direction of motion of the collided particles.
In order to consider the next approximation we should take into account exchange by the virtual gluons that are to be considered as the excitations above the constant (along $x^\pm$) values of $A$. In order to take into account this exchange we embed the obtained $2D$ model back to the $3+1$ space - time and add the conventional action of gluodynamics, that accounts for the processes, in which the virtual gluons participate. Finally, we use the Eikonal approximation for the radiation/absorbtion of multiple virtual gluons by the high energy colliding particles with the Regge kinematics. This approximation allows to calculate the scattering amplitude as the correlator of the two Wilson lines corresponding to the naive trajectories of the colliding particles.  Those three ingredients ($2D$ model, the action for the virtual gluons, and the correlation of the Wilson lines) allows us to derive the effective theory, in which the new auxiliary field is introduced to be identified with the field of the reggeon.

\revision{The given construction is supplemented by the following rule to be used while constructing the perturbation theory. Although in the presence of the reggeon field the leading order classical solution of the equations of motion for $A$ allows to identify those solutions with the reggeon field, the perturbation theory is constructed around the trivial vacuum $A=0$. This rule allows to avoid the overcounting of the reggeons. Alternatively, the coefficient in front of the kinetic term for the reggeon field may be changed. We have considered this on the example of the coupling constants given by Eq. (\ref{naive}). Then the perturbation theory is to be built around the value of $A$ equal to the field of the reggeon $\cal B$. The resulting perturbation theory precisely coincides with the original one made around $A=0$ for the action of Eq. (\ref{Lipat}). This procedure was called finite renormalization of the kinetic reggeon term, and discussed in \cite{Lipatov}.}

As it was  mentioned above, for the simplest (and natural) choice of the coupling constants entering the action of the $2D$ model we reproduce the hermitian version Eq. (\ref{Lipat}) of the action given by L.Lipatov. In Eqs. (210), (218) of \cite{Lipatov} we should substitute the field $A_\pm(v)$ by its anti - hermitian part  $\frac{1}{2}(A_\pm(v)- A^+_\pm(v))$. This transformation guarantees the unitarization of the theory needed for its self - consistency. At the same time this does not change the result of the diagrammatic calculation of the scattering amplitude in the leading approximation,
see Appendix B, and for the details of such calculations see, for example, \cite{Bondarenko:2017vfc}. The same Hermitian version of Lipatov theory has been considered in Eq. (6) of \cite{Nefedov} (see also the works by the same authors \cite{Nefedov2}).

The form of the action of Eq. (\ref{more}) (with the unknown coefficient $\Delta \zeta$) and the form of the action of Eq. (\ref{evenmore}) (with the unknown coefficients $\zeta, \zeta^\prime$) may be considered as improvements of the conventional Lipatov action. The values of the coefficients are to be taken from the comparison of the results obtained using these actions with experiment. We expect, however, that the leading order results will coincide, at least, for the actions of the forms of Eq. (\ref{more}) and Eq.  (\ref{Lipat}). The detailed consideration of this issue is out of the scope of the present paper and will be reported elsewhere.

To conclude, we developed an approach to the derivation of the effective theory of reggeized gluons based on the combination of the effective two - dimensional model of the high energy scattering with the ordinary QCD perturbation theory for the exchange by multiple virtual gluons. The resulting theory represents a version of the effective theory of Lipatov, and it may give its improvement in the sub - leading order.

One of the authors (S.B.)  greatly benefited from numerous discussions with L.Lipatov and A.Prygarin. Both authors are grateful to S.Pozdnyakov for useful discussions.

\section*{Appendix A}

\renewcommand{\theequation}{A.\arabic{equation}}
\setcounter{equation}{0}

In this section we restrict ourselves by the case, when $\zeta,\zeta^\prime$ are given by Eq. (\ref{naive}). We show in Sect. \ref{Lipatov2}, that then the effective action depends on the  anti - Hermitian part of $\cal A$ denoted by $i {\cal B} = \frac{{\cal A}-{\cal A}^+}{2} = i B^a \lambda^a$, where $a = 0,...,8$ marks the generators of the $U(3)$ group.

\revisionZ{Let us consider the two versions of the theory, in which the reggeon field ${\cal B}^\pm$ in the effective action depends/does not depend on $x^\pm$.}

\subsection*{There is the dependence of ${\cal B}^\pm$  on $x^\pm$}

We  have the following expression for the effective action:
\begin{eqnarray}
&& S^\prime_{eff}[A,{\cal B}]  = S^{0}[A]
 \nonumber\\ && \revisionU{-} \frac{2}{g^2}{\rm Tr}\,\int d^4 x \partial_\bot {\cal B}_+(x^+,x^-,x_\bot)\,\partial_\bot {\cal B}_-(x^+,x^-,x_\bot) \\&&
   \revisionU{+}\frac{i}{g^2}\int d^4 x \delta(x^+) {\rm Tr}\,\partial_\bot^2  {\cal B}_-(x) \nonumber\\ &&\Le P\,e^{\imath \,\int_{-\infty}^{\infty}\,d x^+\, \,A^-(x^+,x^-,x_\bot)\,}\,-\,\bar{P}\,e^{-\imath \,\int_{-\infty}^{\infty}\,d x^+\, \,A^-(x^+,x^-,x_\bot)\,}\Ra\, \nonumber\\ &&\nonumber\\&&
 \revisionU{+}\frac{i}{g^2}\int d^4 x \delta(x^-) {\rm Tr}\,\partial_\bot^2 {\cal B}_+(x) \nonumber\\ &&\Le P\,e^{\imath \,\int_{-\infty}^{\infty}\,d x^-\, \,A^+(x^+,x^-,x_\bot)\,}\,-\,\bar{P}\,e^{-\imath \,\int_{-\infty}^{\infty}\,d x^-\, \,A^+(x^+,x^-,x_\bot)\,} \,\Ra\,\nonumber
\end{eqnarray}
Bare reggeon propagator is given by
\begin{eqnarray}
&& \langle {\cal B}_{-}^{a}(x^+,x^-,x_\bot)
  {\cal B}_{+}^{b}(y^+,y^-,y_\bot)\rangle^{(0)}  =  \frac{1}{Z^{(0)}(0)}\, \int D{\cal B}\,\nonumber\\&& {\rm exp}\,\Big(\revisionU{-}i \frac{2}{g^2}{\rm Tr}\,\int d^4 x \partial_\bot {\cal B}_+(x^+,x^-,x_\bot)\,\partial_\bot {\cal B}_-(x^+,x^-,x_\bot)\Big)\nonumber\\&&{\cal B}_{-}^{a}(x^+,x^-,x_\bot)
  {\cal B}_{+}^{b}(y^+,y^-,y_\bot)\nonumber\\&&= -
  \frac{1}{Z^{(0)}(0)}\frac{\delta^2}{\delta J^a_+(x^+,x^-,x_\bot) \delta J^b_-(y^+,y^-,y_\bot)} Z^{(0)}(J)|_{J=0}\nonumber
\end{eqnarray}
with
\begin{eqnarray}
Z^{(0)}(J) & = & \, \int D{\cal B}\, {\rm exp}\,\Big(\revisionU{-}i \frac{2}{g^2}{\rm Tr}\,\int d^4 x \partial_\bot {\cal B}_+(x^+,x^-,x_\bot)\,\nonumber\\&& \partial_\bot {\cal B}_-(x^+,x^-,x_\bot)\nonumber\\&& + i\int d^4 x  {\rm Tr}{\cal B}_{-}(x^+,x^-,x_\bot) J_+(x^+,x^-,x_\bot)\nonumber\\&& + i\int d^4 x {\rm Tr}  {\cal B}_{+}(x^+,x^-,x_\bot) J_-(x^+,x^-,x_\bot)\Big)\nonumber
\end{eqnarray}
The integration over ${\cal B}$ gives
\begin{eqnarray}
Z^{(0)}(J) & = & \, {\rm const} \, {\rm exp}\,\Big(\revisionU{-}i \frac{g^2}{2}{\rm Tr}\,\int d^2  x_\bot dx^+ dx^- \nonumber\\&&{ J}_-(x^+,x^-,x_\bot)\,\frac{1}{\partial^2_\bot} {J}_+(x^+,x^-,x_\bot)\Big)
\end{eqnarray}
We have
\begin{eqnarray}\label{BarProp}
&& \langle {\cal B}_{-}^{a}(x^+,x^-,x_\bot)
  {\cal B}_{+}^{b}(y^+,y^-,y_\bot)\rangle^{(0)}  = \\&&=
  \revisionU{} i \frac{g^2}{8\pi} \,\delta^{ab}\, \delta(x^+-y^+)\delta(x^--y^-)\,{\rm log} \frac{(x_\bot - y_\bot)^2}{\epsilon^2}\nonumber
\end{eqnarray}
where $\epsilon$ is the infrared cutoff. This is precisely the propagator of the reggeon of Eq. (260) from \cite{Lipatov}. \revisionZ{For the components interacting with the virtual gluons we obtain:
\begin{eqnarray}\label{BarProp2}
&&\langle {\cal B}_{-}^{a}(0,x^-,x_\bot)
  {\cal B}_{+}^{b}(y^+,0,y_\bot)\rangle^{(0)}  = \nonumber\\&& =
   i \frac{g^2}{8\pi} \,\delta^{ab}\, \delta(y^+)\delta(x^-)\,{\rm log} \frac{(x_\bot - y_\bot)^2}{\epsilon^2}
\end{eqnarray}
Bare propagator entering Eq. (\ref{DD}) is given by
\begin{eqnarray}\label{BarProp3}
&& \langle \int dx^- {\cal B}_{-}^{a}(x^-,x_\bot)
 \int d y^+ {\cal B}_{+}^{b}(y^+,y_\bot)\rangle^{(0)}  = \nonumber\\&& =
   i \frac{g^2}{8\pi} \,\delta^{ab}\, {\rm log} \frac{(x_\bot - y_\bot)^2}{\epsilon^2}
\end{eqnarray}}

Again, we denote
\begin{eqnarray}
{\Theta}_+(x^-,x_\bot) &=&P\, {\rm exp}\,\Big( i \int A^-(x^+,x^-,\vec{x}_\bot)d x^+ \Big) \nonumber\\&& - \bar{P}\, {\rm exp}\,\Big( -i \int A^-(x^+,x^-,\vec{x}_\bot)d x^+ \Big) \nonumber\\
{\Theta}_-(x^+,x_\bot) &=& P\, {\rm exp}\,\Big( i \int A^+(x^+,x^-,\vec{x}_\bot)dx^- \Big)\nonumber\\&& -\bar{P}\, {\rm exp}\,\Big( -i \int A^+(x^+,x^-,\vec{x}_\bot)d x^- \Big)\nonumber
\end{eqnarray}
and come to the following expression for the partition function:
\begin{eqnarray}
&&Z  =  \, \int D A \, D{\cal B} \,{\rm exp}\,\Big(i S^{(0)}[A] \nonumber\\&& \revisionU{-}i  \frac{2}{g^2}{\rm Tr}\,\int d^4 x \partial_\bot {\cal B}_+(x^+,x^-,x_\bot)\,\partial_\bot {\cal B}_-(x^+,x^-,x_\bot) \\&&
   \revisionU{-}\frac{1}{g^2}\int d^4 x \delta(x^+) {\rm Tr}\,\partial_\bot^2  {\cal B}_-(x)\nonumber\\&&\Le P\,e^{\imath \,\int_{-\infty}^{\infty}\,d x^+\, \,A^-(x^+,x^-,x_\bot)\,}\,-\,\bar{P}\,e^{-\imath \,\int_{-\infty}^{\infty}\,d x^+\, \,A^-(x^+,x^-,x_\bot)\,}\Ra\, \nonumber\\&&
 \revisionU{-}\frac{1}{g^2}\int d^4 x \delta(x^-) {\rm Tr}\,\partial_\bot^2 {\cal B}_+(x) \nonumber\\&& \Le P\,e^{\imath \,\int_{-\infty}^{\infty}\,d x^-\, \,A^+(x^+,x^-,x_\bot)\,}\,-\,\bar{P}\,e^{-\imath \,\int_{-\infty}^{\infty}\,d x^-\, \,A^+(x^+,x^-,x_\bot)\,} \,\Ra\,  \Big)\nonumber\\
 & &=  \, \int D A \,D{\cal B} {\rm exp}\,\Big(i S^{(0)}[A]  \nonumber\\&&\revisionU{-} i  \frac{2}{g^2}{\rm Tr}\,\int d^4 x \partial_\bot \Big({\cal B}_+(x^+,x^-,x_\bot) \revisionU{+} i \frac{\delta(x^+)}{2}\Theta_+(x^-,x_\bot)\Big)\,\nonumber\\&&\partial_\bot \Big({\cal B}_-(x^+,x^-,x_\bot) \revisionU{+}i  \frac{\delta(x^-)}{2}\Theta_-(x^+,x_\bot)\Big)\nonumber \\&&
    \revisionU{-}\frac{i}{2g^2}\int d^4 x \delta(x^+)\delta(x^-) {\rm Tr}\,\partial_\bot \Theta_+(x^-,x_\bot) \, \partial_\bot  \Theta_-(x^+,x_\bot)\Big)\nonumber\\
 && =  \, {\rm const}\,\int D A \,{\rm exp}\,\Big(i S^{(0)}[A]
   \label{LR2} \\&&\revisionU{-} \frac{i}{2g^2}\int d^4 x \delta(x^+)\delta(x^-) {\rm Tr}\,\partial_\bot \Theta_+(x^-,x_\bot) \, \partial_\bot  \Theta_-(x^+,x_\bot)\Big)\nonumber
\end{eqnarray}
This is our starting point for the simplest choice of parameters with the action that consists of the two terms: the usual action of gluodynamics and the action of Eq. (\ref{S+-3}) for the effective two - dimensional model.

\subsection*{No dependence of ${\cal B}^\pm$  on $x^\pm$}

\revisionZ{ Now let us consider the form of the action with the kinetic term for the reggeon field, in which this field does not contain the dependence of ${\cal B}^\pm$  on $x^\pm$. This is the form used traditionally in the works on this subject \cite{EffAct}. In order to obtain it starting from  \eq{Lipat4} we neglect the field $\cal D$. This gives
\beqar\label{AcEf}
&& S_{eff}[A,{\cal B}]  =  S^{0}[A]
 \nonumber\\&&\revisionU{-} \frac{4}{g^2}{\rm Tr}\,\int d^4 x \partial_\bot {\cal B}_+(x^+,x_\bot)\,\partial_\bot {\cal B}_-(x^-,x_\bot)\nonumber \\
 &&\revisionU{+}
\frac{i}{2\,g^2}\int d^4 x\, \partial_{+}\, {\rm Tr}\,\Bigl(
P\,e^{\imath \,\int_{-\infty}^{x^{+}}\,d \bar{x}^+\, \,A^-(\bar{x}^+,x^-,x_\bot)\,}\,\nonumber\\&&-
P\,e^{\imath \,\int_{x^{+}}^{\infty}\,d x^+\, \,A^-(x^+,x^-,x_\bot)\,}\,\nonumber\\
&-&
\bar{P}\,e^{-\imath \,\int_{-\infty}^{x^{+}}\,d \bar{x}^+\, \,A^-(\bar{x}^+,x^-,x_\bot)\,}\,\nonumber\\&&+
\bar{P}\,e^{-\imath \,\int_{x^{+}}^{\infty}\,d \bar{x}^+\, \,A^-(\bar{x}^+,x^-,x_\bot)\,}\,
\Bigr)\, \partial_\bot^2  {\cal B}_-(x^-,x_\bot)-\nonumber\\
&&\revisionU{+}
\frac{i}{2\,g^2}\int d^4 x \,\partial_{-}\, {\rm Tr}\,\Bigl(
P\,e^{\imath \,\int_{-\infty}^{x^{-}}\,d \bar{x}^-\, \,A^+(x^+,\bar{x}^-,x_\bot)\,}\,\nonumber\\&&-\,
P\,e^{\imath \,\int_{x^{-}}^{\infty}\,d \bar{x}^-\, \,A^+(x^+,\bar{x}^-,x_\bot)\,} \,\nonumber \\
&-&
\,\bar{P}\,e^{-\imath \,\int_{-\infty}^{x^{-}}\,d \bar{x}^-\, \,A^+(x^+,\bar{x}^-,x_\bot)\,}\,\nonumber\\&&+\,
\bar{P}\,e^{-\imath \,\int_{x^{-}}^{\infty}\,d \bar{x}^-\, \,A^+(x^+,\bar{x}^-,x_\bot)\,} \,
\Bigr)\,  \partial_\bot^2 {\cal B}_+(x^+,x_\bot)
\eeqar
or
\beqar
&& S_{eff}[A,{\cal B}]  =  S^{0}[A]
 \nonumber\\&& \revisionU{-} \frac{4}{g^2}{\rm Tr}\,\int d^4 x \partial_\bot {\cal B}_+(x^+,x_\bot)\,\partial_\bot {\cal B}_-(x^-,x_\bot)\,\label{AcEf1} \\
 &&\revisionU{+}
\frac{2i}{\,g^2}\int d^4 x\,J_{+}(x^+,x^-,x_\bot)\,\partial_\bot^2  {\cal B}_-(x^-,x_\bot)\,\nonumber \\
&&\revisionU{+}
\frac{2i}{\,g^2}\int d^4 x \,J_{-}(x^+,x^-,x_\bot)\,\partial_\bot^2  {\cal B}_+(x^+,x_\bot)\nonumber
\eeqar
Here
\begin{eqnarray}
&& J_{+}(x^+,x^-,x_\bot) =\frac{1}{4} \partial_{+}\, {\rm Tr}\,\Bigl(
P\,e^{\imath \,\int_{-\infty}^{x^{+}}\,d \bar{x}^+\, \,A^-(\bar{x}^+,x^-,x_\bot)\,}\,\nonumber\\&& -
P\,e^{\imath \,\int_{x^{+}}^{\infty}\,d x^+\, \,A^-(x^+,x^-,x_\bot)\,}\,\nonumber\\
&-&
\bar{P}\,e^{-\imath \,\int_{-\infty}^{x^{+}}\,d \bar{x}^+\, \,A^-(\bar{x}^+,x^-,x_\bot)\,}\,\nonumber\\&&+
\bar{P}\,e^{-\imath \,\int_{x^{+}}^{\infty}\,d \bar{x}^+\, \,A^-(\bar{x}^+,x^-,x_\bot)\,}\,
\Bigr)\,\nonumber\\
&& J_{-}(x^+,x^-,x_\bot) = \frac{1}{4}\partial_{-}\, {\rm Tr}\,\Bigl(
P\,e^{\imath \,\int_{-\infty}^{x^{-}}\,d \bar{x}^-\, \,A^+(x^+,\bar{x}^-,x_\bot)\,}\,\nonumber\\&&-\,
P\,e^{\imath \,\int_{x^{-}}^{\infty}\,d \bar{x}^-\, \,A^+(x^+,\bar{x}^-,x_\bot)\,} \,\nonumber \\
&-&
\,\bar{P}\,e^{-\imath \,\int_{-\infty}^{x^{-}}\,d \bar{x}^-\, \,A^+(x^+,\bar{x}^-,x_\bot)\,}\,\nonumber\\&&+\,
\bar{P}\,e^{-\imath \,\int_{x^{-}}^{\infty}\,d \bar{x}^-\, \,A^+(x^+,\bar{x}^-,x_\bot)\,} \,
\Bigr)\,
\end{eqnarray}
Next, we perform integration over the gluon fields. The classical solutions for the gluon field in the presence of the reggeon $\cal B$ are given by
\beq\label{AcEf4}
A_{cl\,+}(x^+,x^-,x_\bot)\,=-i J_{cl\,+}(x^+,x^-,x_\bot)=\,{\cal B}_+(x^+,x_\bot)
\eeq
and
\beq\label{AcEf5}
A_{cl\,-}(x^+,x^-,x_\bot)\,=-i J_{cl\,-}(x^+,x^-,x_\bot)=\,{\cal B}_-(x^-,x_\bot)\,,
\eeq
We substitute the classical solution back to the action. The resulting expression up to the terms quadratic in the reggeon is
\beqar\label{AcEf6}
&&S^{(2)}_{eff}[A_{cl},\caB]  =  S^{0}[A_{cl}]\,\nonumber\\&&\revisionU{-} \frac{4}{g^2}{\rm Tr}\,\int d^4 x \partial_\bot {\cal B}_+(x^+,x_\bot)\,\partial_\bot {\cal B}_-(x^-,x_\bot)\,-\,\nonumber \\
 &&\revisionU{+}
\frac{2i}{g^2}\int d^4 x\,J_{cl\,+}(x^+,x_\bot)\,\partial_\bot^2  {\cal B}_-(x^-,x_\bot)\,-\,\nonumber \\
&&\revisionU{+}
\frac{2i}{g^2}\int d^4 x \,J_{cl\,-}(x^-,x_\bot)\,\partial_\bot^2  {\cal B}_+(x^+,x_\bot)\nonumber\\
& &\approx  \revisionU{-} \frac{2}{g^2}{\rm Tr}\,\int d^4 x \partial_\bot {\cal B}_+(x^+,x_\bot)\,\partial_\bot {\cal B}_-(x^-,x_\bot)\,
\eeqar
The corresponding (bare) contribution to the reggeon Green function entering Eq. (\ref{DD}) is given by
\begin{eqnarray}
&& \langle \int dx^- {\cal B}_{-}^{a}(x^-,x_\bot)
 \int d y^+ {\cal B}_{+}^{b}(y^+,y_\bot)\rangle^{(0)}  \nonumber\\&&
= \frac{1}{Z^{(0)}(0)}\, \int D{\cal B}\, \nonumber\\&&{\rm exp}\,\Big(\revisionU{-}i \frac{2}{g^2}{\rm Tr}\,\int d^4 x \partial_\bot {\cal B}_+(x^+,x_\bot)\,\partial_\bot {\cal B}_-(x^-,x_\bot)\Big)\nonumber\\&&{\cal B}_{-}^{a}(x^-,x_\bot)
  {\cal B}_{+}^{b}(y^+,y_\bot)\nonumber\\&=& -
  \frac{1}{Z^{(0)}(0)}\frac{\delta^2}{\delta J^a_+(x_\bot) \delta J^b_-(y_\bot)} Z^{(0)}(J)|_{J=0}
\end{eqnarray}
with
\begin{eqnarray}
Z^{(0)}(J) & = & \, \int D{\cal B}\, \nonumber\\&& {\rm exp}\,\Big(\revisionU{-}i \frac{2}{g^2}{\rm Tr}\,\int d^4 x \partial_\bot {\cal B}_+(x^+,x_\bot)\,\partial_\bot {\cal B}_-(x^-,x_\bot)\nonumber\\&& + i\int d^4 x  {\rm Tr}{\cal B}_{-}(x^-,x_\bot) J_+(x_\bot)\delta(x^+)\nonumber\\&& + i\int d^4 x {\rm Tr}  {\cal B}_{+}(x^+,x_\bot) J_-(x_\bot)\delta(x^-)\Big)\nonumber
\end{eqnarray}
The integration over ${\cal B}$ gives
\begin{eqnarray}
Z^{(0)}(J) & = & \, {\rm const} \, {\rm exp}\,\Big(\revisionU{-}i \frac{g^2}{2}{\rm Tr}\,\int d^2  x_\bot  { J}_-(x_\bot)\,\frac{1}{\partial^2_\bot} {J}_+(x_\bot)\Big)\nonumber
\end{eqnarray}
We have
\begin{eqnarray}\label{BarProp8}
&& \langle \int dx^- {\cal B}_{-}^{a}(x^-,x_\bot)
 \int d y^+ {\cal B}_{+}^{b}(y^+,y_\bot)\rangle^{(0)}  \nonumber\\&& =
   i \frac{g^2}{8\pi} \,\delta^{ab}\, {\rm log} \frac{(x_\bot - y_\bot)^2}{\epsilon^2}\nonumber
\end{eqnarray}
This expression coincides with the bare reggeon propagator of Eq. (\ref{BarProp3}).}

\section*{Appendix B}

\renewcommand{\theequation}{B.\arabic{equation}}
\setcounter{equation}{0}

\revisionZ{In this section we propose the alternative formulation of the effective reggeon theory, which coincides with the version discussed in the main text in the leading order. And therefore it should give the same answers for the basic observables in the limit of the multi - regge kinematic. This new version is, possibly, more natural, if we keep the nonzero field $D(x^{-},x^{+}, x_{\bot})$ defined in Eq. (\ref{AddEq1}). Namely, the partition function of Eq. (\ref{Z1}) may be rewritten in the form:
\beqar\label{AC1}
&&Z  =\, \int DA \, {\rm exp}\,\Big(i S^0[A]
   \nonumber\\&&\revisionU{+}{i\zeta_{AB}}{} \int d^4 x \, {\rm Tr}\,\Le\,\partial_{+}\, P\,e^{-A\imath \,\int_{-\infty}^{x^{+}}\,d x^{'+}\, \,A^-(x^{'+},\,0\,,x_\bot)\,}\,\Ra\, \nonumber\\&&\partial_\bot^2
\,\Le\,\partial_{-}\, P\,e^{-B\imath \,\int_{-\infty}^{x^{-}}\,d x^{'-}\, \,A^+(0\,,x^{'-}\,,x_\bot)\,}\,\Ra\,\Big)\,,
\eeqar
This expression may be interpreted as the approximation to the following partition functions
\beqar\label{AC12}
&&Z=\int DA \, {\rm exp}\,\Big(i S^0[A]
  \nonumber\\&& \revisionU{+}{i\zeta_{AB}}{} \int d^4 x\,  {\rm Tr}\,\Le\,\partial_{+}\, P\,e^{-A\imath \,\int_{-\infty}^{x^{+}}\,d x^{'+}\, \,A_{+}(x^{'+},\,x^{-}\,,x_\bot)\,}\,\Ra\, \partial_\bot^2\nonumber\\&&
\,\Le\,\partial_{-}\, P\,e^{-B\imath \,\int_{-\infty}^{x^{-}}\,d x^{'-}\, \,A_{-}(x^{+}\,,x^{'-}\,,x_\bot)\,}\,\Ra\,\Big)\,\nonumber\\&&=\,
{\rm const}\, \int DA D{\cal A}\, {\rm exp}\,\Big(i S_{eff}[A,{\cal A}]\Big)\,,
\eeqar
where
\beqar\label{AC3}
&& S_{eff}[A,{\cal A}]  =  S^{0}[A]
\\&&\revisionU{+}{\zeta_{AB}} {\rm Tr}\,\int d^4 x \,\partial_\bot {\cal A}_+^A(x^+,x^-,x_\bot)\,\partial_\bot {\cal A}_-^B(x^+,x^-,x_\bot)\label{SL}\,-\,\nonumber\\&&
  \revisionU{+}\zeta_{AB}{\rm Tr}\, \int d^4 x \,\partial_{x^+} \Le P\,e^{-A\imath \,\int_{-\infty}^{x^+}\,d \bar{x}^+\, \,A_{+}(\bar{x}^+,x^-,x_\bot)\,}\Ra\,
	\nonumber\\&&\partial_\bot^2 {\cal A}_-^B(x^{+},\,x^-,\,x_\bot)\nonumber\\&&
  \revisionU{+}\zeta_{AB}{\rm Tr}\, \int d^4 x\, \partial_{x^-} \,\Le P\,e^{-B\imath \,\int_{-\infty}^{x^-}\,d \bar{x}^-\, \,A_{-}(x^+,\bar{x}^-, x_\bot)}\,\Ra\,
\nonumber\\&&\partial_\bot^2 {\cal A}_+^A(x^+\,,x^{-}\,,x_\bot)\nonumber
\eeqar
In this case, performing the same steps as in the main text of the paper, we obtain instead \eq{SL} the following four - dimension expression for the effective action:
\beqar\label{AC4}
&\,& S_{eff}[A,{\cal B}] =  S^{0}[A]\\&&
  \revisionU{-}\frac{2}{g^2}{\rm Tr}\,\int d^4 x \partial_\bot {\cal B}_+(x^+,x^-,x_\bot)\,\partial_\bot {\cal B}_-(x^+,x^-,x_\bot)\,-\,\nonumber\\
 &&\revisionU{+}
\frac{i}{g^2}\int d^4 x\, \partial_{+}\, {\rm Tr}\,\Bigl(
P\,e^{\imath \,\int_{-\infty}^{x^{+}}\,d \bar{x}^+\, \,A^-(\bar{x}^+,x^-,x_\bot)\,}\,\nonumber\\&&-
\bar{P}\,e^{-\imath \,\int_{-\infty}^{x^{+}}\,d \bar{x}^+\, \,A^-(\bar{x}^+,x^-,x_\bot)\,}\Bigr)\,
\partial_\bot^2  {\cal B}_-(x^{+},\,x^-,\,x_\bot)-\nonumber\\
&&\revisionU{+}
\frac{i}{g^2}\int d^4 x \,\partial_{-}\, {\rm Tr}\,\Bigl(
P\,e^{\imath \,\int_{-\infty}^{x^{-}}\,d \bar{x}^-\, \,A^+(x^+,\bar{x}^-,x_\bot)\,}\,\nonumber\\&&-\,
\,\bar{P}\,e^{-\imath \,\int_{-\infty}^{x^{-}}\,d \bar{x}^-\, \,A^+(x^+,\bar{x}^-,x_\bot)\,}\,\Bigr)\,
\partial_\bot^2 {\cal B}_+(x^+,\,x^{-},\,x_\bot)\,\nonumber
\eeqar
The effective action written in the main text and, correspondingly, the original action of Lipatov, may be obtained from \eq{AC4} when in the reggeon field the dependence of ${\cal B}^\pm$  on $x^\pm$ is absent. The suppression of this dependence occurs for the kinematical reasons due to the smallness of the ratio $|t/s|$. Using the action of the form of Eq. \eq{AC4} we may calculate the sub - leading corrections to the physical observables. To what extent those corrections are relevant for the dynamics in the case of the multi - Regge kinematics is not completely clear. However, the action of the form of \eq{AC4} may appear to be more useful from the point of view of practical calculations, even if we consider the leading order contributions only . }

\end{document}